\documentclass{aa}

\usepackage{graphicx}
\usepackage{txfonts}
\usepackage{hyperref}
\usepackage{xcolor}
\defcitealias{2020A&A...641A..51F}{F20}

\newcommand{\ndet}{1561}  % number of total detections with eROSITA
\newcommand{\ndetmem}{655}  % number of detected members with eROSITA
\newcommand{\ndetmemtotal}{668}  % number of detected members with eROSITA, Chandra, XMM
\newcommand{\ndetmemrot}{433}  % number of detected rotators

\begin{document}

\title{A star-by-star correspondence between X-ray activity and rotation in the young open cluster NGC\,2516 with eROSITA}

   \author{D. J. Fritzewski
          \inst{1,2}
          \and
          S. A. Barnes
          \inst{1,3}
          \and
          S. Ok
          \inst{1,4}
          \and
          G. Lamer
          \inst{1}
          \and
                  A. Schwope
          \inst{1}
          }

   \institute{Leibniz Institute for Astrophysics Potsdam (AIP), An der Sternwarte         16, 14482 Potsdam, Germany%\\
              \and
              Institute of Astronomy, KU Leuven, Celestijnenlaan 200D, 3001 Leuven, Belgium\\
              \email{dario.fritzewski@kuleuven.be}
              \and
              Space Science Institute, 4750 Walnut St., Boulder, CO 80301, USA
              \and
Department of Astronomy \& Space Sciences, Faculty of Science, University of Ege, 35100 Bornova, Izmir, Turkey
             }

   \date{}
% \abstract{}{}{}{}{}
% 5 {} token are mandatory

  \abstract
    {The coronal soft X-ray emission of cool stars, especially when taken in combination with their measured rotation periods,  offers insights into their levels of magnetic activity and related transitions.}
   {We study the X-ray properties of low-mass (FGKM-type) members of the open cluster NGC\,2516 to explicate their detailed dependencies on mass and rotation.}
   {We analysed the pointed SRG/eROSITA satellite observations of NGC\,2516 obtained during the calibration and performance verification phase of the mission. We found \ndet{} X-ray sources within the field of view and related 1007 of them to their optical stellar counterparts, including \ndetmem{} members of NGC\,2516 (\ndetmemrot{} of which have measured rotation periods). We combined these detections with auxiliary optical data
   to facilitate their interpretation. Furthermore, we extracted X-ray spectra for all sources and fit two-component \texttt{APEC} models to them. To aid the analysis, we grouped stars with similar mass and rotational properties together, which allowed us to investigate the influence of rotation on various X-ray properties.}
   {The colour-activity diagram (CAD) of NGC\,2516 displays a general increase in the fractional X-ray luminosity with spectral type change from F through G and K to M-type. However, the behaviour of K-type stars, representing the ones that best sample the fast-to-slow rotational transition, is more complex, with both increased and decreased X-ray emission relative to G-type stars for fast and slow rotators, respectively. The rotation-activity diagram is analogous, with an identifiable desaturated group of X-ray emitters that corresponds to stars in the rotational gap between the fast and slow rotator sequences. We prefer to describe the normalised X-ray emission for all cluster stars as declining logarithmically with Rossby number over those using broken power laws. Coronal temperatures appear to be largely independent of mass or rotation. The coronal abundances are significantly sub-solar for most stars, as shown in prior works, but near-solar for both the least-active stars in our sample and a few of the very active stars as well.}
   {The wealth of X-ray detections and rotation periods for stars in NGC\,2516 enables a detailed view of the connection between rotation and X-ray activity in a homogeneous and coeval cluster sample of young stars.}

    \titlerunning{Correspondence between X-ray activity-rotation in NGC\,2516}

   \maketitle
%
%-------------------------------------------------------------------

\section{Introduction}
Cool solar-type stars emit soft X-rays from their hot coronae, indicative of stellar magnetic activity. The strength of this emission depends on stellar rotation and convection, both believed to be key ingredients of stellar dynamos \citep{1955ApJ...121..491P, 1962AnAp...25...18S}. The depth and strength of convection in cool stars is known to be a function of mass (or its measured observational proxy:\ colour), while the preferred measure of rotation is the rotation period. A specific point of interest for this study is to examine the existence (or otherwise) of a star-by-star correspondence in X-rays for a known rotational transition between fast and slow rotation. As such, fairly comprehensive information about both X-rays and rotation is necessary for a large and homogeneous sample of stars -- one that we endeavour to provide in this study through our examination of the young (Pleiades-age) open cluster NGC\,2516.

Open clusters are particularly useful targets for such studies because their member stars share properties such as age, composition, and space motion, while also covering a wide range of spectral types or masses. Thus, they offer snapshots of stellar evolution across different masses and provide clear pictures of relations that are obscured in studies of field stars; the latter including stars of different ages, rotation rates, and levels of activity even if the spectral type is kept fixed.

The decline of chromospheric activity with age was first noted in \cite{1963ApJ...138..832W} and followed up on in \cite{1964ApJ...140.1401W}. A parallelism between the decay of rotation  and chromospheric activity, both proposed to decline as power laws with age, seems to have initially been recognised in \cite{1972ApJ...171..565S}, while the connection between rotation and activity for field stars of differing ages was first explored carefully in \cite{1984ApJ...279..763N}.  In the meantime, \cite{1981ApJ...248..279P} showed that this connection between rotation and activity is also valid for coronal X-ray activity. Based on their observations, they suggested that the corona is not heated directly by the dissipation of rotational energy, but rather by the intermediary of the magnetic field. Parker Solar Probe observations \citep{2016SSRv..204....7F}, especially those described in the study by \cite{2023Natur.618..252B}, have strengthened long-standing expectations that this heating does indeed arise from magnetic reconnection and a wave-induced dissipation of energy.

A key insight, first enunciated in \cite{1984ApJ...279..763N}, is that the dimensionality of the rotation-activity-colour problem can be reduced by rescaling the rotation period with the convective turn-over timescale to obtain a quantity called the Rossby number \citep[see also][]{1993SoPh..145..207D}, considerably decreasing the scatter in the (rescaled) rotation-activity diagram\footnote{In keeping with prior usage we continue to call it the rotation-activity diagram despite the fact that the ordinate really is the Rossby number rather than the rotation period.}. Later works have typically followed this ansatz and analysed aspects of this connection, finding a power-law dependence of the X-ray activity on the Rossby number (e.g \citealt{2003A&A...397..147P}, \citealt{2011ApJ...743...48W}). However, this relationship has not been found to be informative for fast-rotating stars. Stars with small Rossby numbers (indicating fast rotation) have activity levels independent of the rotation rate \citep{1984A&A...133..117V,1987ApJ...321..958V}. This part of the relationship is called the saturated regime, while stars following the power-law dependence are correspondingly described as being in the unsaturated regime.

Because of their proximity, field stars have been the preferred targets of most prior works. For such stars, their interpretation in terms of the Rossby number is almost unavoidable when rotation periods were actually available because no clear relationships are evident otherwise. However, a Rossby number-centric interpretation also disguises features of the mass-dependence in coeval populations. In addition, studies of the connection between X-ray activity and rotation were often hindered by the unavailability of the rotation period, necessitating the use of the more ambiguous rotation velocity, ${\rm v} \sin i$. Here, $\rm{v}$ represents the star's equatorial rotation velocity and $i$ is the generally unknown angle of inclination. This situation is actively being ameliorated with rotation period measurements, both from ground-based data, and more recently, from space data \citep[e.g.][]{2018A&A...616A..16L, 2022AJ....164..137K, 2023A&A...674A..20D}.

Cluster stars offer distinct advantages over field stars, as alluded to earlier. As a result, a number of nearby clusters and associations have been studied in X-rays \citep[e.g][]{2022A&A...661A..40S}, including the nearby Pleiades \citep[e.g.][]{1993A&A...277..114S}, Hyades, and Praesepe clusters. In fact, \cite{2022ApJ...931...45N, 2024ApJ...962...12N}  recently performed a comparable study on the connection between rotation and X-ray activity for the two $\sim${}650\,Myr old open clusters Hyades and Praesepe. We aim to provide a similarly comprehensive analysis for the young ($\sim${}150\,Myr) main sequence open cluster NGC\,2516 with the present study.

Although the open cluster \object{NGC\,2516} studied here was previously observed extensively in X-rays, rotation periods for FGK stars became available only recently \citep{2020A&A...641A..51F, 2020ApJ...903...99H, 2021AJ....162..197B}. Consequently, the rotation-activity connection was not the main focus of earlier works \citep[e.g.][]{1997MNRAS.287..350J, 2006A&A...450..993P}. The same can be said about other open clusters for which large sets of rotation periods only recently started to become available (e.g. \citealt{2016AJ....152..113R}, \citealt{2016ApJ...823...16B}, \citealt{2016ApJ...822...47D}, \citealt{2020A&A...644A..16G}, \citealt{2020A&A...641A..51F}, \citealt{2021A&A...652A..60F}). Therefore, dedicated studies of the connection between X-ray activity and rotation in main sequence open clusters are not numerous. The following list gives, to our knowledge, a complete overview
(Pleiades: \citealt{1995PASP..107..211P}, \citealt{1999A&A...341..751M};
$\alpha$\,Per: \citealt{1996A&A...305..785R};
IC\,2391: \citealt{1996ApJS..106..489P};
NGC\,6475: \citealt{1997MNRAS.292..252J};
NGC\,2547: \citealt{2006MNRAS.367..781J};
NGC\,2451: \citealt{2004A&A...418..539H};
M\,34: \citealt{2012A&A...546A.117G};
M\,35: \citealt{2013A&A...556A..14G};
M\,37: \citealt{2015ApJ...809..161N};
Hyades: \citealt{2020A&A...640A..66F}; and
Hyades and Praesepe: \citealt{2022ApJ...931...45N, 2024ApJ...962...12N})

Here, we study the coronal properties of stars in the young, southern open cluster NGC\,2516, observed as part of the calibration and performance verification segment (CalPV) of the extended ROentgen Survey and Imaging Telescope Array (eROSITA, \citealt{predehl+21}; for more information on observations see Sect.~\ref{sec:obs}) to calibrate the telescope's boresight. NGC\,2516 has served the same purpose for the earlier XMM-\emph{Newton} and \emph{Chandra} missions because of its abundance of young stars, combined with their extent on the sky, allowing the detectors to be well-sampled across the field of view (FoV).

The first X-ray observations of NGC\,2516 were obtained with the ROSAT mission and analysed in \cite{1996A&A...312..818D}, \cite{1997MNRAS.287..350J}, and \cite{2000A&A...357..909M}. In addition, several subsequent studies utilising the various calibration observations from \emph{Chandra} and XMM-\emph{Newton} have been published: \cite{1999STIN...0012868W, 2001ApJ...547L.141H, 2003ApJ...588.1009D} (\emph{Chandra}) and \cite{2001A&A...365L.259S, 2006A&A...450..993P, 2006A&A...456..977M} (XMM-\emph{Newton}). NGC\,2516 is an ideal target for investigating the connection between rotation and stellar activity. Not only is this cluster very rich, with the eROSITA observations providing many X-ray detections among cluster members, but recently the number of measured rotation periods in this open cluster has increased significantly. The initial set of rotation periods consisted mainly of M\,dwarfs \citep{2007MNRAS.377..741I}. In our previous work \citep{2020A&A...641A..51F}, we provided rotation periods for GKM stars. Shortly thereafter, \cite{2020ApJ...903...99H} and \cite{2021AJ....162..197B} used data from the Transiting Exoplanet Survey Satellite (TESS) to further enhance the set of known rotators with F and G stars.

NGC\,2516 is a southern twin of the Pleiades. Both clusters are very rich and nearly coeval, with an age of $\sim${}150\,Myr \citep[hereafter F20]{1993A&AS...98..477M, 2002AJ....123..290S,2020A&A...641A..51F}. Due to these properties NGC\,2516 is the best-studied southern open cluster, the target of many photometric (e.g. \citealt{1997MNRAS.287..350J,2001A&A...375..863J,2002AJ....123..290S,2006A&A...453..101L,2024A&A...686A.142L}) and spectroscopic studies (e.g. \citealt{2002ApJ...576..950T,2016A&A...586A..52J,2018MNRAS.475.1609B}). More detailed information on NGC\,2516 as an open cluster (membership, rotation, etc.) can be found in our previous work \citepalias{2020A&A...641A..51F}.

In \citetalias{2020A&A...641A..51F}, we also provided an inclusive membership list based on multi-colour photometry, \emph{Gaia} DR2 astrometry and parallaxes, and radial velocities. This list is the basis for the analysis of the GKM cluster member rotation periods in \citetalias{2020A&A...641A..51F}. Furthermore, we also provided the first rotation-X-ray activity diagram for NGC\,2516, based on the XMM-\emph{Newton} observations of \cite{2006A&A...450..993P}. In this work, we extend those efforts and analyse the rotation-activity connection in greater detail.

This paper is structured as follows. In Sect.~\ref{sec:obs}, we present our observations and data reduction. This is followed by a search for the optical counterparts of the X-ray detections in Sect.~\ref{sec:counterp}. In Sect.~\ref{sec:adddata}, we present the auxiliary data used in our analysis. In Sects.~\ref{sec:CAD} and \ref{sec:CADrot}, we analyse the connection between X-ray activity, rotation, and stellar mass in NGC\,2516. It is followed by Sect.~\ref{sec:specanalysis}, where we investigate the connection of spectral properties with other stellar parameters. Finally, we present our conclusions in Sect. 8.

\section{Observations}
\label{sec:obs}
As the major component
of the Spectrum-Roentgen-Gamma (SRG) satellite, eROSITA \citep{predehl+21} was launched  on 13 July 2019  \citep{sunyaev+21}. During the first months of the mission, a series of calibration and performance verification (CalPV) observations were carried out. NGC\,2516 was observed twice to calibrate the boresights and plate scales of the eROSITA telescopes. The combined exposure time was 140\,ks (see Table \ref{tab:ero_obs}). For comparison, the total exposure of NGC\,2516 in the four eROSITA all-sky surveys completed to date is 1.4\,ks. Data from both CalPV observations are publicly available as part of the eROSITA early data release.

\begin{table}
    \caption{Observations of NGC\,2516 during the eROSITA CalPV phase.}
    \label{tab:ero_obs}
    \begin{tabular}{llll}
        \hline
        \hline
        ObsID & Start date   &  Good time & Telescope units\\
        \hline
        700018 & 2019-10-06  & 56250 s   & TM5..TM7 \\
        700019 & 2019-10-31  & 77686 s   & TM1..TM7 \\
        \hline
    \end{tabular}
\end{table}

\subsection{Data preparation}

For this work, we processed the data with the eROSITA Science Analysis Software System (\texttt{eSASS}, \citealt{brunner+22}) version \texttt{c020}. This pipeline provides calibrated event files and images in standard energy bands. Before merging the two observations, we applied astrometric corrections to both datasets.

\begin{figure}
    \includegraphics[width=\columnwidth]{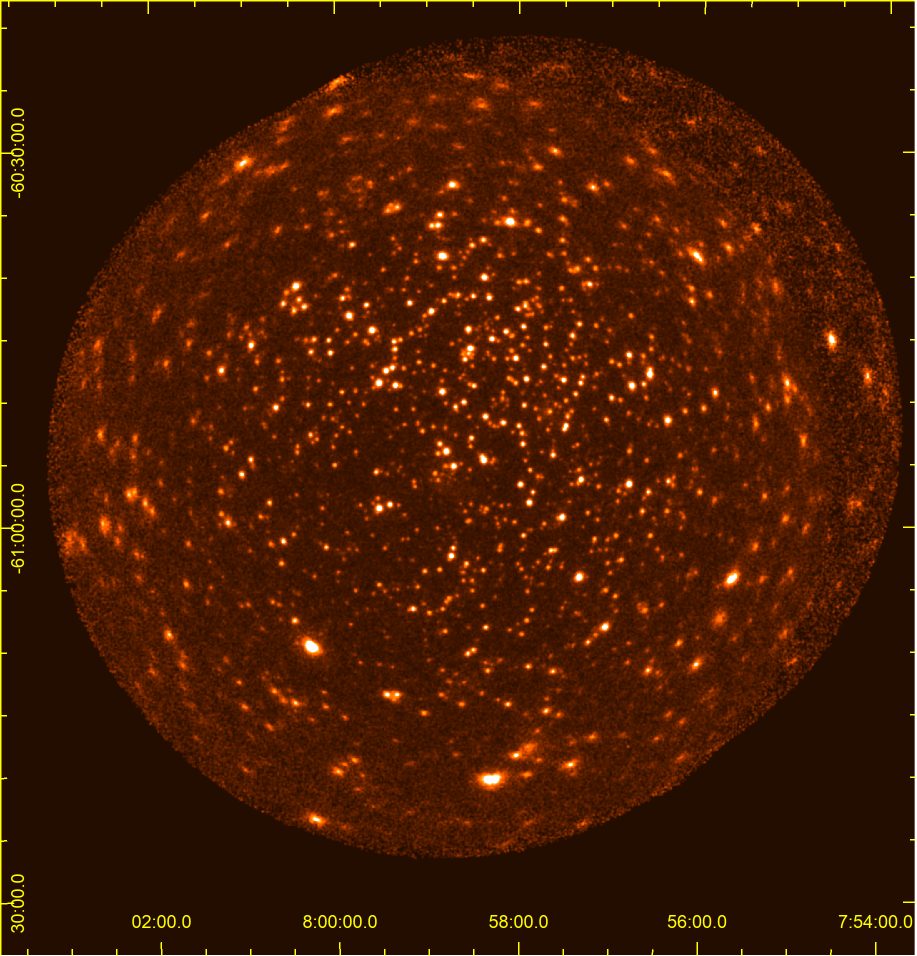}
    \caption{
    X-ray image of the NGC\,2516 region in the $0.2-2.3$\,keV band, obtained by stacking the two eROSITA pointed observations of NGC\,2516.
    The image is approximately one degree across and has been exposure corrected and slightly smoothed.
    }
    \label{fig:image}
\end{figure}

To determine the remaining boresight offsets, we generated preliminary \texttt{eSASS} source lists for both observations and assigned the corresponding {\em Gaia} counterparts \citep{2021A&A...649A...2L} within a 12\arcsec{} matching radius. To find the  optimal 3D rotation matrix, $R_\mathrm{opt}$, minimising the differences between X-ray positions and {\em Gaia} reference positions, we applied the method described in \cite{markley14} using singular value decomposition (SVD). For each of the two observations, the matrix $R_\mathrm{opt}$ defines an offset in the equatorial coordinates, $\alpha$, $\delta$, and in the roll angle, $\phi$. The resulting roll angle corrections of about $0.02\deg$ amount to sub-arcsecond positional shifts even at the largest off-axis angles and are neglected. The following corrections were applied to each event:
\begin{align}
\alpha_\mathrm{corr} &= \alpha_\mathrm{uncorr} + \Delta\alpha/ \cos\delta,\\
\delta_\mathrm{corr} &= \alpha_\mathrm{uncorr} + \Delta\delta,
\end{align}
with $ \Delta\alpha=-2.087\arcsec{}$, $\Delta\delta=+0.052\arcsec{}$ for observation 700018 and $ \Delta\alpha=-2.835\arcsec{}$, $\Delta\delta=+4.685\arcsec{}$ for observation 700019. The corrected event lists were then merged to create a stacked image in the energy band $0.2-2.3$\,keV (see Fig. \ref{fig:image}). An exposure map of the combined image was calculated using the \texttt{eSASS} task \texttt{expmap}. This step takes into account the actual exposure times as well as any corrections due to telescope vignetting, detector dead-time, or cameras that were not used during the observation.

\subsection{Source detection and extraction}

The X-ray source catalogue created with the \texttt{eSASS} detection pipeline is based on the stacked image in the $0.2-2.3$\,keV band.  For the \texttt{eSASS} detection tasks, we used parameter settings very similar to the those described in \cite{brunner+22}. However, since the deep observations of NGC\,2516 are characterised by a very high surface density of point sources, we optimised the detection pipeline for the deblending and characterisation of point sources. To achieve the best possible separation, we applied the smallest box size ($20\arcsec\times 20 \arcsec$) in the \texttt{erbox} task and performed no iterative rebinning of the image. We set the PSF fitting task \texttt{ermldet} to fit point source models only and allowed multi-PSF fitting to fit up to four input sources simultaneously within a 1\arcmin{} radius, while enabling the modelling to split up each input source into two sources. In total, the detection pipeline yielded \ndet{} X-ray sources in the eROSITA FoV. Since the detection pipeline applies the exposure corrections contained in the exposure map, all source count rates in the FoV are equivalent to the source count rate detected at the centre of the FoV with all cameras switched on. The PSF fitting also corrects the count rates for the source flux beyond the source fitting region.

For each source, we calculated the X-ray fluxes in the eROSITA detection band ($0.2-2.3$\,keV) by dividing the model count rates by a global energy conversion factor (ECF). To calculate the ECF, we simulated a template X-ray spectrum using \texttt{XSPEC} \citep{arnaud+96, dorman+01} and the relevant eROSITA calibration files. From these inputs, we were able to determine the model flux and count rate in the $0.2-2.3$\,keV band. We opted to use a two-temperature \texttt{APEC} model as our spectral model. For the galactic absorption value towards NGC\,2516 ($N_\mathrm{H}= 8 \cdot 10^{20} \mathrm{cm}^{-2} $) and the chemical abundance value ($Z=0.3$) of the stellar atmosphere, we used the values determined by \cite{2006A&A...450..993P}. We set the two \texttt{APEC} temperatures to $k_\mathrm{B}T_1 = 0.3\,\mathrm{keV}$ and $k_\mathrm{B} T_2 = 1.0\,\mathrm{keV}$ with equal normalisation. These values are the typical results from the eROSITA spectral analysis of the NGC\,2516 cluster members (see below). We note that the application of this ECF yields the unabsorbed flux of the stars. Although it is only true for cluster members, we applied the resulting $\mathrm{ECF} = 7.95 \cdot 10^{11} \mathrm{\,cm^2\,erg^{-1}}$ to all the sources in the catalogue. To estimate the systematic error in the energy to flux conversion due to deviations of the X-ray spectra from the canonical spectrum, we also calculated a set of ECFs for the best fit spectra of the 20 brightest cluster members. The resulting mean ECF is $ 7.80 \cdot 10^{11} \mathrm{\,cm^2\,erg^{-1}} $ with a standard deviation of $ 0.2 \cdot 10^{11} \mathrm{\,cm^2\,erg^{-1}} $, suggesting a systematic $1\sigma$ error of 2.6\,\%.

\begin{table}
    \caption{Overview on the table columns of the online Table of all \ndet{} X-ray detections.}
    \label{tab:raw_data}
    \resizebox{\columnwidth}{!}{
    \begin{tabular}{lll}
        \hline
        \hline
        Name & Unit & Description\\
        \hline
        SRC\_ID & - & Identifier\\
        RA & $\deg$ & Right ascension\\
        Dec & $\deg$ & Declination\\
        RADEC\_err & $\deg$ & Positional uncertainty\\
        Rate & s$^{-1}$ & Count rate \\
        Rate\_err & s$^{-1}$  & Uncertainty in count rate \\
        Flux & $10^{-14}$\,erg\,cm$^{-2}$\,s$^{-1}$ & X-ray flux (using ECF for NGC\,2516)\\
        Flux\_err& $10^{-14}$\,erg\,cm$^{-2}$\,s$^{-1}$ & Uncertainty in X-ray flux\\
        \hline
    \end{tabular}
}
\tablefoot{The full table is available at the CDS.}
\end{table}

\subsection{Spectral fitting}

In addition to the integrated X-ray flux, we wished to analyse other data products from eROSITA. Therefore, we extracted the background corrected light curves, spectra and related files, response matrices, and ancillary responses with the \texttt{srctool} task in \texttt{eSASS}.

\label{sec:spec_fit}

We performed a spectral analysis of the all sources\footnote{The main focus of our work is on the cool star members of NGC\,2516. As we establish the connection between X-ray sources and members  later in this work, we applied the spectral fit to all sources. Subsequently, non-stellar sources were removed and we could disregard their obviously incorrect spectral fits.} with \texttt{XSPEC} version 12.12.0 \citep{arnaud+96, dorman+01}. The spectral fitting was applied to the energy range $0.2 - 2.3$\,keV. The nominal energy range of eROSITA reaches up to 10 keV. However, due to a sharp drop of the effective area at 2.3 keV, the spectra of softer sources such as coronal emitters are dominated by instrumental background beyond this energy. Since there are not enough photons to fit the binned spectra of the faintest sources using the $\chi^2$ statistics for optimisation, we used a binning with one count per bin and utilised the \texttt{cstat} statistics \citep{cash79,kaastra17}.

\begin{figure}
    \includegraphics[width=\columnwidth]{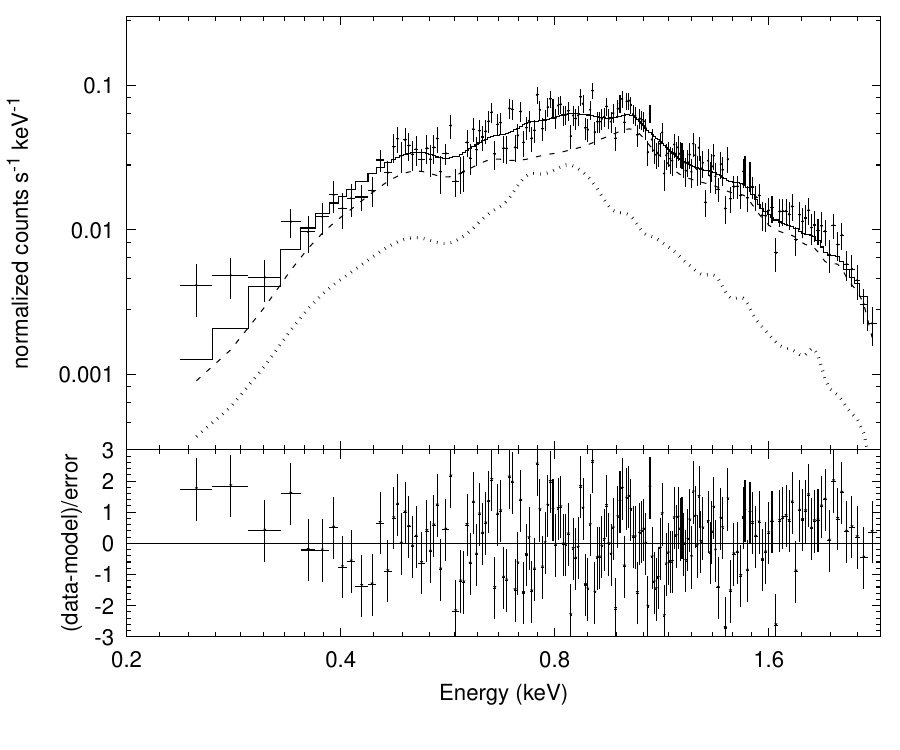}
    \caption{Example of an eROSITA X-ray spectrum, namely of the cluster member ID\,5. The spectrum is fitted with a two-component \texttt{APEC} model (solid line) with respective temperatures of $k_\mathrm{B} T_1 = 0.68$\,keV (dotted line) and $k_\mathrm{B} T_2 = 1.24$\,keV (dashed line). The lower panel shows the normalised residuals.}
    \label{f:xspec}
\end{figure}

We started by modelling the sources with a single thermal emission \texttt{APEC} \citep{smith+01} model with galactic absorption (\texttt{TBABS}) \citep{wilms+00}. To apply this model globally to all sources, we used a Python module of \texttt{XSPEC} named \texttt{PyXspec} \citep{gordon+21} and performed the fit. This model causes strong residual features  around 1\,keV, in particular, for bright objects with high signal-to-noise ratios (i.e. counts greater than $\sim$500). To improve the fit results, we added another \texttt{APEC} model to the initial model, leading to a significantly  improved description of the spectra (see Fig.~\ref{f:xspec} for an example). In the \texttt{APEC} models, we used the \texttt{wilms} \citep{wilms+00} metal abundance tables and fixed the redshift parameters to zero. We linked the abundance parameters in both \texttt{APEC} models to each other and released the abundance and temperature parameters to perform the fit in \texttt{XSPEC}.

Since there are many different objects in the field, we initially released the \texttt{TBABS} parameter to investigate the total hydrogen column density ($N_{\rm H}$) distribution. The \texttt{TBABS} distribution scatters around $N_{\rm H}=0.08\times10^{22}\,\mathrm{cm}^2$ for cluster members. This is consistent with the $N_\mathrm{H}$ value towards NGC\,2516 resulting from the XMM-\emph{Newton} observation previously analysed by \cite{2006A&A...450..993P}. For the final analysis, we therefore fixed the \texttt{TBABS} parameter to $N_\mathrm{H}=0.08\times10^{22}\,\mathrm{cm}^2$. The results of the final spectral fits are listed in Table~\ref{tab:spectra}.

\begin{table}
    \caption{Overview of the online table of measured spectral properties for optically matched sources.}
    \label{tab:spectra}
    \resizebox{\columnwidth}{!}{
    \begin{tabular}{lll}
        \hline
        \hline
        Name & Unit & Description\\
        \hline
        SRC\_ID & - & Identifier\\
        RA & $\deg$ & Right ascension\\
        Dec & $\deg$ & Declination\\
        RADEC\_ERR & $\deg$ & Uncertainty in position\\
        APEC1\_kT & keV & First \texttt{APEC} component\\
        APEC1\_kT\_err\_l & keV & Lower uncert. of first \texttt{APEC} comp.\\
        APEC1\_kT\_err\_u & keV & Upper uncert. of first \texttt{APEC} comp.\\
        APEC2\_kT & keV & Second \texttt{APEC} component\\
        APEC2\_kT\_err\_l & keV & Lower uncert. of second \texttt{APEC} comp.\\
        APEC2\_kT\_err\_u & keV & Upper uncert. of second \texttt{APEC} comp.\\
        mean\_kT & keV & Weighted mean of the two \texttt{APEC} comp.\\
        mean\_kT\_err & keV & Uncertainty of the mean\_kT\\
        abundance & - & Abundance of \texttt{APEC} fit\\
        abundance\_err & - & Uncert. in abundance of \texttt{APEC} fit\\
        Probability & - & Null hypothesis probability of the fit \\
        \hline
    \end{tabular}
}
\tablefoot{Column density fixed to the cluster value of $N_\mathrm{H}=0.08\times 10^{22}$\,cm$^{-2}$. Full table is available at the CDS.}
\end{table}

To use the fitted temperatures in a meaningful way, we only included results with well-constrained uncertainties. To this end, we removed all results where the uncertainty was equal to or greater than the measured value. This left us with 408 values for the primary component and 536 values for the secondary, hotter component. We obtained a true two component fit for 305 sources. Only these sources were then used in our later analysis. We also extracted light curves in the two energy bands $0.5 - 2$\,keV and $2 - 10$\,keV. However, they were not analysed systematically for all sources in this work.

\section{Optical counterparts}
\label{sec:counterp}
\subsection{Source matching}
\label{sec:match}
As the focus of this study is on stellar X-ray emission at the cluster distance, we matched the extracted source positions to a cleaned \emph{Gaia} sample of stars in the direction of NGC\,2516. We selected only \emph{Gaia} sources with a significant parallax, $\varpi/\sigma_{\varpi}>3$, as recommended by \cite{2022A&A...661A...6S}. This criterion excludes distant sources because they do not have accurately measured parallaxes. We considered only matches within 7\arcsec{} because, for larger distances, the contamination fraction\footnote{In hindsight, this decision can also be justified by the achieved astrometric precision.} becomes too large \citep{2022A&A...661A...6S}.

In total, we found 1157 \emph{Gaia} sources matching 1007 X-ray sources to within 7\arcsec{}. The mean surface density of the filtered Gaia catalogue around the eROSITA sources is $1.55\,\mathrm{arcmin}^{-2}$, leading to an expected number of 87 spurious matches (see Appendix~\ref{app:matches} for more details). For each source with multiple optical counterparts, we selected the most plausible solution. Typically, we chose the closest source, since, in most cases, this was also a main sequence star at the cluster distance; this choice was very likely to ensure an optical counterpart. Some sources were matched to two cluster members with similar matching distances. In such cases, we selected the optically brighter star. However, we indicate in Table~\ref{tab:optical} that there are two optical sources that could be associated with the one X-ray source. All optical counterparts to the X-ray detections are provided in Table~\ref{tab:optical}, together with their related properties. There were 554 sources left over that were not matched to any stellar source and they  are not discussed  further in this paper. However, we do include their positions and count rates in Table~\ref{tab:raw_data} to allow them to be used for other purposes.

\begin{figure}
    \includegraphics[width=\columnwidth]{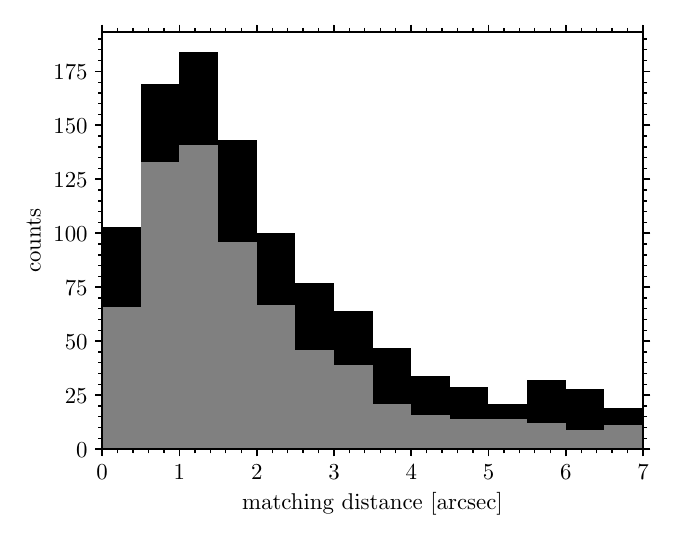}
    \caption{Histogram of the separation distances (black) between all eROSITA sources and the corresponding best-match \emph{Gaia} sources. The equivalent match with the subsample of cluster members is shown in grey. The median of both distributions is well short of 2\arcsec{} (see text).
    }
    \label{fig:mdist}
\end{figure}

Figure~\ref{fig:mdist} shows (in black) the distribution of separations between all optical counterparts to the X-ray source. The majority of eROSITA sources have a stellar counterpart within 2\arcsec{}, with the median distance being only 1.74\arcsec{}, demonstrating the accurate astrometry achievable with eROSITA. For cluster members (grey in Fig.~\ref{fig:mdist}) these separations are even slightly smaller, with only 11\,\% of the distances exceeding 4\arcsec{}. Hence, we are confident that we have identified most of the observed cluster members correctly.

\subsection{Bolometric flux}
After successfully matching the X-ray sources with the optical counterparts, we proceeded to calculate the bolometric flux for each target.
Firstly, we estimated the effective temperature ($T_\mathrm{eff}$) for each star based on the relation from \cite{2019MNRAS.482.2770C} via the dereddened\footnote{$E(B-V)=0.11$ \citep{2002AJ....123..290S}, with $R(g-K_s)=2.678$ (\citealt{2014MNRAS.444..392C},  http://skymapper.anu.edu.au/filter-transformations/)} $(g-K_s)_0$ colour, using the Skymapper $g$ and 2MASS $K_s$ magnitudes. For stars with insufficient photometry in either of the surveys, we switched to the relation from \cite{2021MNRAS.507.2684C}, which has a slightly larger uncertainty in $T_\mathrm{eff}$ but relies only on \emph{Gaia} photometry \citep{2021A&A...649A...3R}. Neither of these relations is valid for the least-massive stars in our survey. Therefore, for all stars with $T_\mathrm{eff}<4500$\,K or a colour outside of the validity range of \cite{2019MNRAS.482.2770C} and \cite{2021MNRAS.507.2684C}, we used the more accurate low-mass star relations from \cite{2015ApJ...804...64M}. To apply these relations to the $(V-J)_0$ colour, we took the colour from \cite{2001A&A...375..863J}. We had to estimate $V$ from \emph{Gaia} $G$ for certain stars missing in the photometry of \cite{2001A&A...375..863J}. For this transformation, we used an polynomial fit to the cluster members with both magnitudes measured.

With the effective temperatures at hand, we calculated the bolometric correction for each star. For stars with $T_\mathrm{eff}$ based on \cite{2019MNRAS.482.2770C} or \cite{2021MNRAS.507.2684C}, we used the bolometric correction as given by \cite{2018A&A...616A...8A} and applied it to the dereddened \emph{Gaia} $G_0$ magnitude to obtain the bolometric flux\footnote{Stars with $T_\mathrm{eff}>8000$\,K are outside the scope of the relation given by \cite{2018A&A...616A...8A}. In our sample 29 stars are removed for this reason. As we are interested in the cool stars, this restriction does not influence our science case.}, $F_\mathrm{bol}$. For stars with lower $T_\mathrm{eff}$ estimates, we used the dedicated bolometric correction from \cite{2015ApJ...804...64M} because we noticed that the \emph{Gaia} bolometric corrections \citep{2018A&A...616A...8A} overestimate the bolometric flux among the least massive stars in our observations.

From the bolometric flux and the X-ray flux, we calculated the fractional X-ray luminosity ($\log L_\mathrm{X}/L_\mathrm{bol}$, as listed in Table~\ref{tab:optical}). In the uncertainty of $F_\mathrm{bol}$ (as part of the uncertainty of the fractional X-ray luminosity), we included only the contributions of the bolometric corrections provided in \cite{2018A&A...616A...8A} and \cite{2015ApJ...804...64M} because the uncertainties in the \emph{Gaia}-derived parameters and effective temperature ($\sigma_{T_\mathrm{eff}}=33\,\mathrm{K} - 64\,\mathrm{K}$, depending on the applied relationship) are small in comparison with the uncertainties of the bolometric corrections.

\begin{table}
    \caption{Overview of the online table of the optical counterparts, derived X-ray, and related properties.}
\label{tab:optical}
\resizebox{\columnwidth}{!}{
\begin{tabular}{lll}
    \hline
    \hline
    Name & Unit & Description\\
    \hline
    SRC\_ID & - & Identifier\\
    RA & $\deg$ & eROSITA right ascension\\
    Dec & $\deg$ & eROSITA declination\\
    designation & - & \emph{Gaia} DR3 designation\\
    ra & $\deg$ & \emph{Gaia} DR3 right ascension\\
    dec & $\deg$ & \emph{Gaia} DR3 declination\\
    Teff & K & Effective Temperature\\
    phot\_g\_mean\_mag & mag & $G$ From \emph{Gaia} DR3\\
    g\_rp0 & mag & $(G-G_\mathrm{RP})_0$ colour from \emph{Gaia} DR3\\
    logLXLbol & - & Fractional X-ray luminosity\\
    err\_logLXLbol & - & Uncertainty of $\log L_\mathrm{X}/L_\mathrm{bol}$\\
    member & - & Indicate cluster membership\\
    Per & d & Rotation period\\
    Ro\_Rosun & - & Solar-scaled Rossby number\\
    multiple & - & Indicates whether two sources are within 7$\arcsec$\\
    \hline
\end{tabular}
}
\tablefoot{The full table is available at the CDS.}
\end{table}

\section{Auxiliary data and prior X-ray observations of NGC\,2516}
\label{sec:adddata}
A number of comparisons and additional steps are necessary before the eROSITA data can be effectively interpreted. These include a comparison with prior X-ray observations of NGC\,2516 and consideration of the membership, binarity, and rotation period information.

\subsection{Comparison with \emph{Chandra} and XMM-\emph{Newton} data}
\label{sec:comp}
The whole set of \emph{Chandra} boresight calibration observations were analysed in \cite{2003ApJ...588.1009D}, while the corresponding XMM-\emph{Newton} calibrations were analysed in \cite{2006A&A...450..993P}. This allows us to limit our comparison to one publication of the \emph{Chandra} dataset and one for the XMM-\emph{Newton} dataset.

As a calibration cluster, NGC\,2516 was observed by \emph{Chandra} with all imaging instruments in fields roughly centred on the open cluster. Due to the differing FoVs of these instruments, the exposure times are inhomogeneous across the cluster members.
The stars in \cite{2003ApJ...588.1009D} have been observed for between 10\,ks and 49\,ks in the individual observations with the different instruments of \emph{Chandra}. For the ACIS detector \cite{2003ApJ...588.1009D} chose the $0.3-8$\,keV band. The central field was observed five times. \cite{2003ApJ...588.1009D} provided only converted luminosities with an assumed distance. For this work, we calculated the flux values and afterwards the relative luminosity with the same \emph{Gaia}-based bolometric flux as for the eROSITA observations.

NGC\,2516 was also observed by XMM-\emph{Newton} for boresight corrections on six occasions, with exposure times between 9\,ks and 21\,ks. \cite{2006A&A...450..993P} chose the $0.3-7.9$\,keV band of the EPIC detector for their analysis (in contrast to the $0.2-2.3$\,keV in the eROSITA band). To increase comparability, we recalculated $\log L_\mathrm{X}/L_\mathrm{bol}$ with the bolometric flux from \emph{Gaia} and did not use the values provided by \cite{2006A&A...450..993P}.

\begin{figure}
    \includegraphics[width=\columnwidth]{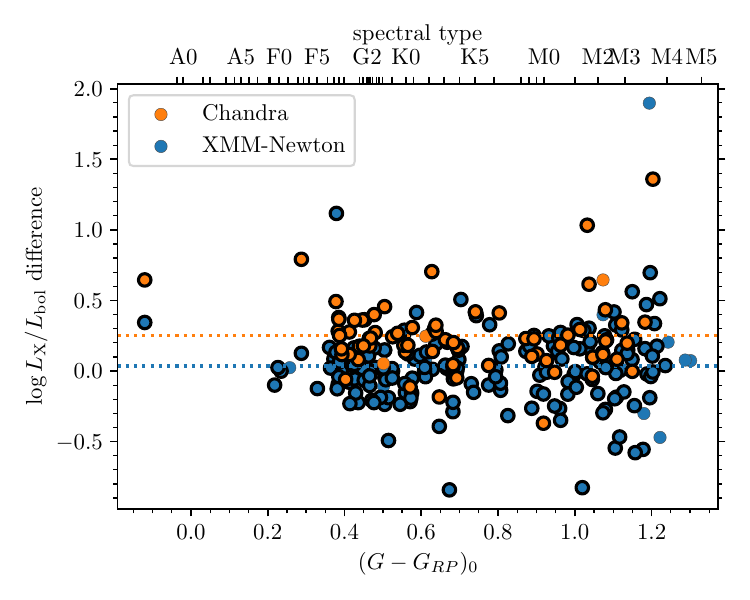}
    \caption{Differences in $\log L_\mathrm{X}/L_\mathrm{bol}$ between eROSITA and \emph{Chandra} (orange), and between eROSITA and XMM-\emph{Newton} (blue). The differences are in the sense of ($\mathrm{Other} -\mathrm{eROSITA})$. The dotted lines show the mean offsets between the datasets. These offset values were used to adjust the prior data to the eROSITA level when necessary for cross-comparisons. Data points outlined in black indicate cluster members.}
    \label{fig:compdiff}
\end{figure}

In Fig.~\ref{fig:compdiff}, we compare the relative X-ray luminosities of the three missions\footnote{The spectral types used in this figure and throughout the paper are based on \cite{2013ApJS..208....9P} (\url{http://www.pas.rochester.edu/~emamajek/EEM\_dwarf\_UBVIJHK\_colors\_Teff.txt)}}.
For both \emph{Chandra} and XMM-\emph{Newton}, we find that the reported fluxes are similar to those of our eROSITA observations. For \emph{Chandra}, we find a mean difference of 0.25\,dex with a large scatter ($\sigma=0.26$\,dex), independent of the stellar colour. We find the detections with the lowest flux to have the largest offset. However, brighter sources also show offsets, likely explained by flares in the \emph{Chandra} data.

Correspondingly, when we integrate all the data together, we lower the obtained fractional X-ray luminosities of the \emph{Chandra} observations by that amount to make the data comparable. The mean difference between the eROSITA and XMM-\emph{Newton} data is 0.036\,dex ($\sigma=0.28$\,dex), effectively compatible without any offset (Fig.~\ref{fig:compdiff}, blue).

Since the eROSITA observations cover a very large region around NGC\,2516 in comparison with prior \emph{Chandra} and XMM-\emph{Newton} data, we would expect to find most of the sources of prior observations in our dataset as well. Indeed, we find that only 17 sources (including 13 cluster members) that are present in the prior observations are absent from the eROSITA data. Of these, 13 come from the \emph{Chandra} dataset. We appended these 17 sources to our data for further analysis. We suspect that these additional sources are blended in the eROSITA observations due to the lower resolution as compared with \emph{Chandra}.

\subsection{Membership}
To interpret the eROSITA X-ray detections, we used the membership list from our prior work \citepalias{2020A&A...641A..51F} as the basis and extended it further with the membership determinations of \cite{2020A&A...640A...1C} and \cite{2021A&A...645A..84M}. This updated joint membership includes 1165 stars in an area somewhat larger than the eROSITA FoV.

\begin{figure*}
    \sidecaption
    \includegraphics[width=12cm]{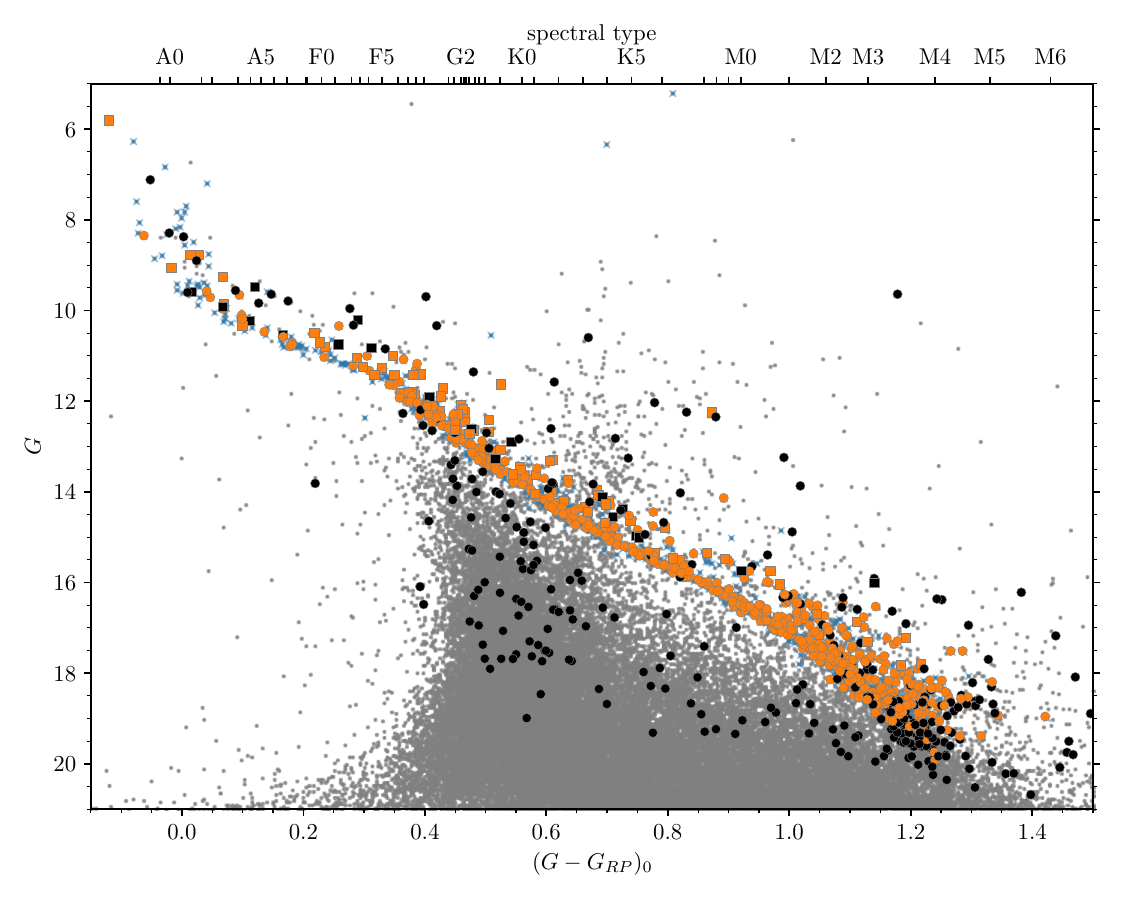}
    \caption{Colour-magnitude diagram of the field observed by eROSITA. Large symbols indicate stars with eROSITA X-ray detections. Of these, members of NGC\,2516 are displayed in orange, while non-members detected in X-rays are in black. NGC\,2516 cluster members undetected with eROSITA are marked with small blue crosses, while field stars (undetected in X-rays) are shown with small grey dots. Potential (member) binaries from radial velocity \citep{2018MNRAS.475.1609B, 2020A&A...641A..51F}, astrometry (\emph{Gaia} RUWE), or photometry (above the main sequence in \emph{Gaia} photometry) are additionally marked with squares. Certain stars on the cluster upper main sequence (clearly photometric members) are marked as non-members because they do not meet the formal (and restrictive) membership criteria set up for our predominantly cool star sample (see text).
    The few cluster members well above the cluster sequence are on it in other photometry, indicating an issue with their \emph{Gaia} photometry.}
        \label{fig:CMD}
\end{figure*}

The majority of the X-ray sources matched with a cool star in the FoV are cluster members (67\,\%). Consequently, the membership analysis seems to have primarily removed obvious non-members (including 140 foreground stars, but see below) from our sample. Among our detected eROSITA sources, we found \ndetmem{} cluster members. In combination with the \emph{Chandra} and XMM-\emph{Newton} data, we have a set of \ndetmemtotal{} cluster members detected in X-rays. The colour-magnitude diagram (CMD) in Fig.~\ref{fig:CMD} demonstrates the richness of X-ray detections on the cluster sequence. In particular, in combination with the wealth of low-mass star rotation periods (see below), this dataset is unique in size for a single open cluster.

As apparent from Fig.~\ref{fig:CMD}, many cluster members are not detected in X-rays. The non-detections on the upper main sequence are expected, since only a fraction of intermediate mass stars emit X-rays \citep{2007A&A...475..677S}. Of the 250 non-detections in the eROSITA FoV, we find the majority to be off-axis, where the detection limit is higher. About 80 low-mass members are not detected in the central part of the FoV. Despite our de-blending, we suspect that many of these stars could not be detected due to the high source density near the cluster core. The non-detected members are distributed along the entire low-mass regime and are not limited to a certain spectral type. Similarly, their rotation periods are equally distributed between 1 and 10\,d. Therefore, we can conclude that the non-detections do not bias our further analysis for the FGK stars. We  return to this issue for M-type cluster members in Sect.~\ref{nuanced}.

Among the stars of the upper main sequence ($(G_\mathrm{BP}-G_\mathrm{RP})_0<0.3$, A-type stars) of NGC\,2516 (but also some at lower masses), some X-ray detections on the cluster sequence are marked as non-members in Fig.~\ref{fig:CMD}. These stars, despite being located at the right distance, exhibit kinematics that are not in agreement with the cluster and with the tidal tails. Many of these stars, nominally classified as non-members (despite being obvious photometric members) have large RUWE values (often $\mathrm{RUWE}>3$) in \emph{Gaia} DR3, indicating that the astrometric solution is likely inaccurate for these very bright stars. Because of this tension, we have excluded them from the nominal membership list. We note that this decision is irrelevant to our science as our main focus is on the cool star members of NGC\,2516.

\subsection{Rotation periods}
Apart from convection, the crucial driver of the stellar dynamo and, therefore, of coronal emission is stellar rotation. Thus, rotation periods are key to characterising and understanding coronal X-ray emission. A number of recent studies have measured rotation periods for stars in NGC\,2516. These are \cite{2007MNRAS.377..741I}, \cite{2020A&A...641A..51F}, \cite{2020ApJ...903...99H}, and \cite{2021AJ....162..197B}. We  collected the rotation periods from these studies and conjoined them into a single dataset. Because each of these studies had differing sensitivity and targeted somewhat different mass regimes and also sky positions, the overlap between them is actually quite small. Nevertheless, when there is overlap the periods of stars in the various studies typically agree within the uncertainties \citep{2021AJ....162..197B}.

\begin{figure}
    \includegraphics[width=\columnwidth]{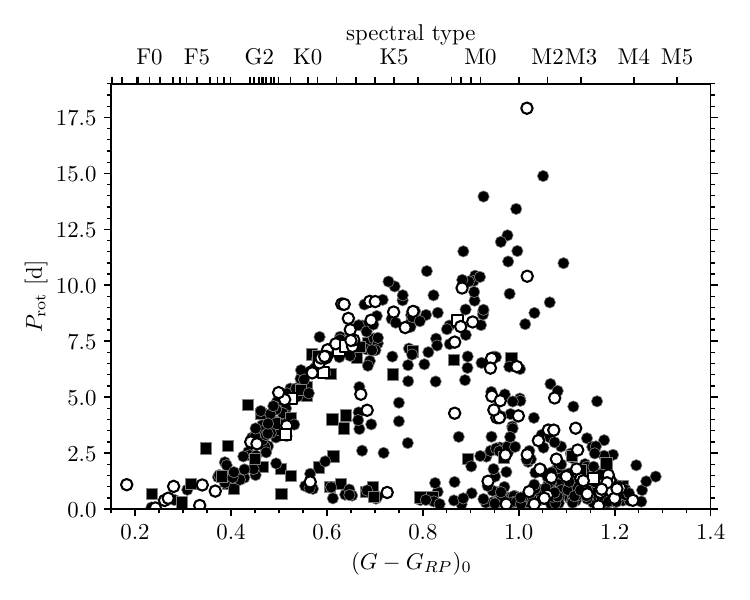}
    \caption{Colour-period diagram for NGC\,2516 cluster members. Filled symbols represent cluster members also detected in X-rays, while open symbols indicate rotators within the eROSITA FoV not detected in X-rays. Rotation periods are from \citetalias{2020A&A...641A..51F} and additional studies in the literature (see text). The diagram is well populated in all relevant regions, allowing us to investigate the X-ray properties of stars with a very wide range of rotation and mass/spectral type without a bias against certain types.
    Binaries are indicated with squares whereas stars without signs of binarity are marked with circles.
     }
\label{fig:CPD}
\end{figure}

We find that of the \ndetmemtotal{} likely cluster X-ray sources, \ndetmemrot{} (65\,\%) have measured rotation periods.
In Fig.~\ref{fig:CPD}, we show the colour-period diagram (CPD) for this combined subset of rotators with measured X-ray emission in NGC\,2516. For completeness, we also show rotators in the eROSITA FoV without X-ray detections. All regions of the CPD, including the fast rotators, slow rotators, and both the slow- and fast-rotating M\,dwarfs are well represented here.

\subsection{Binarity}
Stellar interactions in binaries can influence both the rotation rates of the components and the stellar activity levels. This influence is likely restricted to close binaries. Correspondingly, we did not find a significant influence of binarity (defined broadly) on stellar activity in a similar study of the 300\,Myr-old open cluster \citep{2021A&A...656A.103F}. Nevertheless, we include a brief analysis of the binarity status of the studied cluster members here. In addition, as seen in Sect.~\ref{sec:match} a certain number of X-ray sources can be matched to two (likely) binary cluster members. The measured properties therefore contain signatures of both components, which could bias our results.

NGC\,2516 is a well-studied open cluster that was the subject of prior radial velocity studies. For our work, we use the radial velocity binaries published in \cite{2018MNRAS.475.1609B} and in \citetalias{2020A&A...641A..51F}. \cite{2021AJ....162..197B} have also identified many photometric binaries in NGC\,2516. In addition to these stars, we have included some stars not in the sample of \cite{2021AJ....162..197B}, based on their offset relative to the cluster main sequence. Finally, we used the \emph{Gaia} DR3 reduced unit weight error (RUWE) and classified all stars with $\rm{RUWE}>1.2$ as potential astrometric binaries. RUWE is not a direct measure of astrometric binarity; only a goodness of fit parameter. However, the perturbations in the astrometric solution can arise from close, unseen companions through photocentre shifts \citep{2020MNRAS.496.1922B}. In total, we find 79 potential binaries among our X-ray selected sources. In the following, we mark all potential binaries in the figures with distinct symbols (squares) to make it easier to identify and locate them.

\section{Connections between rotation, activity, and mass}
\label{sec:CAD}
\subsection{Rotation-activity relationship}
The observed X-ray activity of low-mass stars emerges from their coronae, themselves almost certainly heated via magnetic reconnection \citep[e.g.][]{2023Natur.618..252B}. The associated magnetic fields in turn are believed to be generated by stellar dynamos, driven by rotation and convection \citep[e.g.][]{1955ApJ...122..293P}. Hence, it is natural to investigate the X-ray activity in concert with stellar rotation and convection. The standard way of connecting these quantities,, which we have also used here, is via the dimensionless Rossby number, $\mathrm{Ro} = P_\mathrm{rot}/\tau_c$, where $\tau_c$ is the convective turnover timescale in stars. For this work, we use $\tau_c$ from \cite{2010ApJ...721..675B}. To enhance comparability with work that prefers other sources for $\tau_c$, we scale all Rossby numbers to the Solar value, using the solar rotation period, $P_{\mathrm{rot},\sun}=26.09$\,d \citep{1996ApJ...466..384D}. Such normalised Rossby numbers should largely be compatible across different publications because the individual numerical values can largely be brought into agreement using a single numerical scaling factor (e.g. \citealt{2010ApJ...721..675B} and Fig.~6 in \citealt{2011ApJ...741...54C}).

Regardless of the choice of the particular convective turnover timescale, it transpires that the usage of $\mathrm{Ro}$ considerably reduces the mass-dependent scatter in the rotation-activity diagram, informing us that it is likely that there is an underlying physical basis for its usage. This approach goes back to \cite{1984ApJ...279..763N} who investigated chromospheric activity measurements. With the availability of many soft X-ray observations from ROSAT the analysis of stellar activity in the context of rotation could also be expanded to coronal emission. The first rotation-X-ray activity diagrams in the modern sense were constructed for the two young open clusters IC\,2391 \citep{1996ApJS..106..489P} and $\alpha$\,Persei \citep{1996A&A...305..785R}. We analyse the rotation-activity relation for NGC\,2516 in the same tradition.

\subsubsection{Overall appearance}
\label{sec:rotact}
\begin{figure}
    \includegraphics[width=\columnwidth]{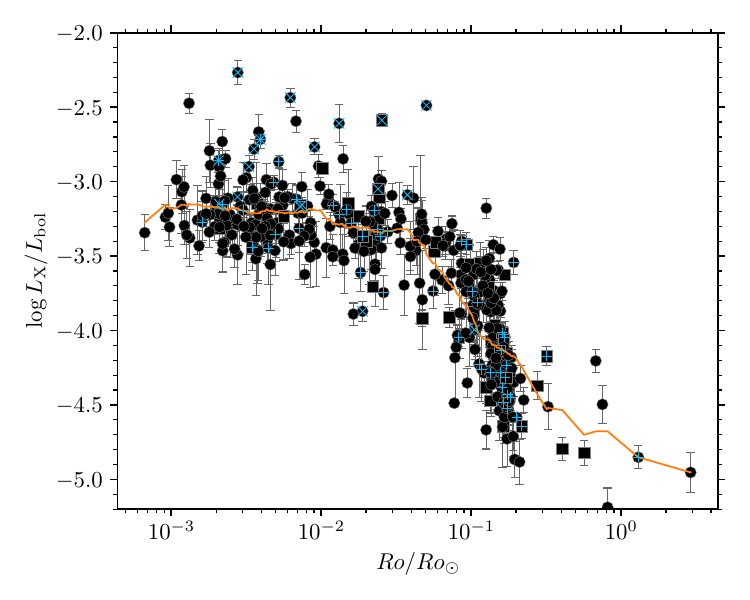}
    \caption{Rotation-activity diagram for NGC\,2516 with the fractional X-ray luminosity $\log L_\mathrm{X}/L_\mathrm{bol}$ plotted against the solar-scaled Rossby number ($Ro/Ro_\sun$). The orange line shows a moving average for a window of $\Delta\log Ro/Ro_\sun=0.2$. Members with identified flares in their light curves are marked with blue x-shaped symbols, and members with two potential optical counterparts with blue pluses. The two canonical regimes (of saturated and unsaturated stars) are obvious. However, a gradual decline of the X-ray activity in the less densely populated region between the flat, saturated regime and the steeply declining unsaturated regime is also noticeable. Stars rightward of $Ro/Ro_\sun = 0.25$ are of early F-type, and are only displayed here for completeness.
    }
    \label{fig:rotact}
\end{figure}

Figure~\ref{fig:rotact} displays the fractional X-ray luminosity ($\log L_\mathrm{X}/L_\mathrm{bol}$) against the solar-scaled Rossby number for cluster members of NGC\,2516. The overall appearance has the well-known shape of a saturated region at low Rossby number where the normalised X-ray flux is largely independent of rotation rate, and a correlated, unsaturated regime towards higher Rossby number, where the flux declines rapidly with slower rotation.

This diagram is typically described as a two-segmented function of Rossby number and fitted with a broken power-law \citep[e.g.][]{2003A&A...397..147P, 2011ApJ...743...48W} with a constant activity level in the saturated regime and a power-law dependence in the unsaturated regime. However, our observations, as presented in Fig.~\ref{fig:rotact}, suggest that, although there is a sharp transition from the saturated to the unsaturated regime (as noted by earlier authors), there is also apparently a small decline within the saturated regime as stars move from $Ro/Ro_\sun \lesssim 0.005$ to $Ro/Ro_\sun \sim 0.05$, as can be seen in the behaviour of the moving average line in Fig.~\ref{fig:rotact}. We note that \cite{2020A&A...638A..20M} have also reported a non-constant saturation level for field M\,dwarfs which could be related to our observations. Our stars, of course, encompass not only M-type, but also F-, G-, and K-types, and they are all members of the one cluster, NGC\,2516.

We find the fastest rotators ($Ro/Ro_\sun<0.03$) to be located mostly at a constant level of activity, but towards slightly larger Rossby numbers ($Ro/Ro_\sun\gtrsim0.03$) there is a noticeable drop as seem from the running mean shown in Fig.~\ref{fig:rotact}. Near the classical saturation limit ($Ro/Ro_\sun\gtrsim0.1$ for $\tau_c$ from  \cite{2010ApJ...721..675B}, \citealt{2021A&A...656A.103F})\footnote{For comparison, we provide a version of Fig.~\ref{fig:rotact} with $\tau_c$ from \cite{2011ApJ...741...54C} in the Appendix Fig~\ref{fig:rotactCS}.} the activity level seems to have already declined by $\sim$0.5\,dex compared to the saturated fast rotators. We conclude that the rotation-activity relation is therefore likely not simply a two-segmented function of Rossby number but could have additional intermediate steps as proposed by \citep[e.g.][]{2003ApJ...586L.145B} if the evolution is not actually one of continuous decline.

In the \cite{2003ApJ...586L.145B} picture, the rotation-activity relation consists of a saturated regime ($Ro<0.03$), a sparsely populated `desaturated' regime ($0.03\lesssim Ro/Ro_\sun\lesssim 0.1$) called the `gap' region in that publication, and the unsaturated regime of spun down, classical slow rotators. The desaturated regime has previously not been noticed in X-ray data\footnote{We note that in our analysis of the chromospheric activity of stars in NGC\,3532 \citep{2021A&A...656A.103F}, we did find a similar shape in the rotation activity diagram.} because it was relatively sparsely populated in open clusters. For example in our previous work on NGC\,2516 \citepalias{2020A&A...641A..51F}, we analysed the rotation-activity connection using literature data available at that point, and a gradual decline is not obvious in those data. The new eROSITA data, by virtue of their quantity and quality, have enabled us to observe this feature, and we are optimistic that further data derived from current large surveys (eROSITA, TESS, \emph{Gaia}, LAMOST, SDSS, etc.) will soon enable the community to investigate this issue in greater detail.  Before we can analyse the current data in more detail, we wish to investigate possible outliers to the sequences and the mass dependence of the X-ray activity.

\subsubsection{Outliers}
Before proceeding further, we need to discuss outliers. For instance, there are a few outliers to the left of the unsaturated sequence (or, equivalently, below the saturated stars) in the rotation activity diagram (Fig.~\ref{fig:rotact}). These stars could have true rotation periods larger than the measured values, a circumstance which would put them onto the unsaturated sequence. The original measurements could have been affected by a half-period alias, which can arise for instance from spot groups on opposite hemispheres of a star. As the number of outliers is small relative to the overall sample, we prefer not to modify any previously measured rotation periods accordingly, and leave any related investigations to future work.

Among the fastest rotators in our sample (those with small Rossby numbers) are some stars that exceed the saturation level even within the uncertainties. Some of these stars are flagged as having multiple optical counterparts (marked with pluses in the figure). Thus, it is likely that these stars have an increased X-ray luminosity due to blending. For the remaining stars, we have checked the X-ray light curves and found that most of them are flaring variables (marked with \emph{X}-es in the figure). The general trend is that the most X-ray luminous stars exhibit strong flares in the X-ray time series. Stars just slightly above the saturation limit tend to show only a low level of variability (and are not marked as flaring). We conclude that stars above the saturation limit are those affected by X-ray variability or flux contamination from additional sources.

One very peculiar star can be found well above the sequence at $Ro/Ro_\sun\approx 0.05$. This star (ID 5) shows a strong flaring signature with a factor ten higher flux compared to the quiescent level. A detailed analysis of the light curves with respect to flare properties such as frequency and energy shall be presented and discussed elsewhere.

To the right of the unsaturated sequence, with $Ro/Ro_\sun > 0.2$ in Fig.~\ref{fig:rotact}, we find a number of stars with very large calculated Rossby numbers. These stars are all actually among the bluest stars in our sample and are expected to have extremely thin convection zones. In this regime the usual correlation between X-ray activity and Rossby number for late-type stars might well break down (Sect.~\ref{sec:kraft}), although these stars may still have (shallow) convective envelopes, rotate, and emit (coronal) X-rays. The offset to the right does not depend on the particular choice of $\tau_c$ as seen from Fig.~\ref{fig:rotactCS}. Hence, we exclude all stars with $Ro/Ro_\sun>0.2$ from the further analysis below.

\subsection{Dependence on colour (mass)}
\subsubsection{Colour-activity diagram}

\begin{figure*}
    \includegraphics[width=\textwidth]{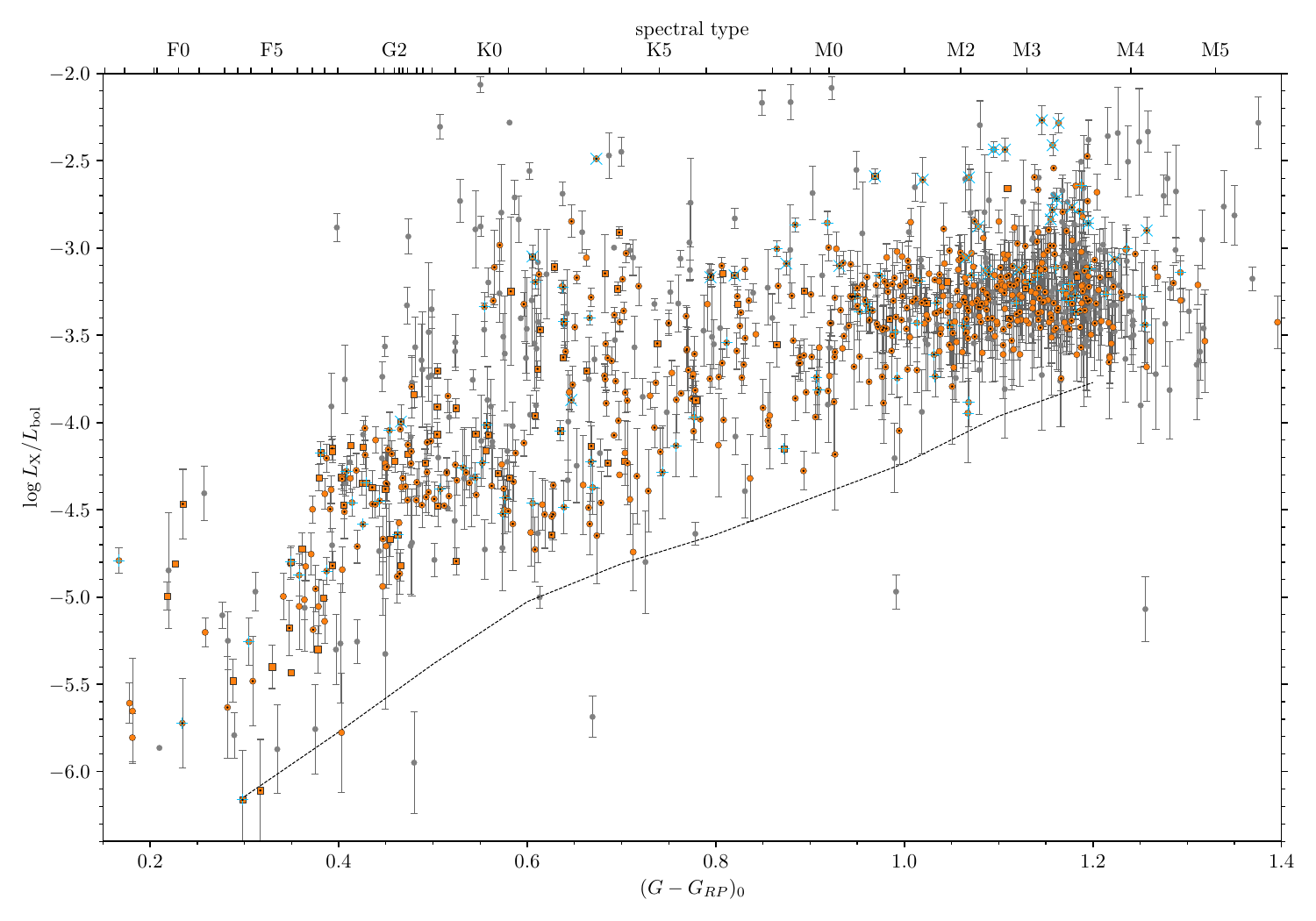}
    \caption{Colour-activity diagram based on the combined set of all observations for stars in the NGC\,2516 field.
    We show the fractional X-ray luminosity ($\log L_\mathrm{X}/L_\mathrm{bol}$) against \emph{Gaia} $(G-G_\mathrm{RP})_0$ colour for cluster members (orange) and cluster non-members (small, grey symbols). Cluster binaries are indicated with orange squares. Members with measured rotation periods are additionally marked with a small black dot, members with identified flares in their light curves with an x, and members with two potential optical counterparts with pluses. Overall, the lower mass stars have X-ray activity exceeding that of higher mass stars by orders of magnitude, reaching the saturation limit for M-type and many K-type stars. The dashed line indicates an estimate of the detection limit for cluster members.
}
    \label{fig:CAD}
\end{figure*}

Further conclusions about the activity can be drawn from the dependence of X-rays at fixed age on stellar mass (for which colour serves as a good proxy). This is only possible because we investigate the rotation and activity in a coeval population of stars, namely, an open cluster. In Fig.~\ref{fig:CAD}, we show this dependence using $(G-G_\mathrm{RP})_0$ as our proxy for stellar mass by constructing what we, for obvious reasons, call a colour-activity diagram (CAD). This particular colour choice confers the benefit of using measured and precise \emph{Gaia} colours for the largest possible sample, including for the reddest M-type dwarfs\footnote{Of course, it is possible to construct the equivalent diagram in other colours or even $\rm{T}_{\rm{eff}}$, as listed in Table 4, at the cost of completeness. We prefer $(G-G_\mathrm{RP})_0$ over $\rm{T}_{\rm{eff}}$ because the former is a directly measured quantity, keeping the CAD observational.}.

The members of NGC\,2516 follow a roughly diagonal sequence stretching from quite low activity levels ($\log L_\mathrm{X}/L_\mathrm{bol} \sim -6$) for the higher mass stars (spectral type F) towards saturated activity in the lower mass stars ($\log L_\mathrm{X}/L_\mathrm{bol} \sim -3$). The potential cluster binaries (marked with squares) follow the same sequence as the single members. In this particular figure, we also display cluster non-members (small grey symbols). These non-members can be observed both on and off the cluster-occupied regions at both higher and lower activity levels. In particular among G-type stars, we find a significant number of non-members above the observed sequence and even above the saturation limit. These stars are mostly likely to be very young field stars or short period binary systems that happen to be located in the NGC\,2516 observing field.

We see that stars of a given spectral type (except for late-G to K-type stars) have a relatively well-defined X-ray emission rate, with a spread of only about 0.5\,dex (a factor of three). Stars of a common age and mass, and therefore rotation rate, have similar activity levels. This is promising from the point of view of associating stars of a given spectral type with a particular age purely from measured X-ray activity levels. Among the K-type stars, however, we found a wide spread of activity levels in the range $-4.8\le L_\mathrm{X}/L_\mathrm{bol}\le -2.8, $  which we find to be the effect of the wide distribution in rotation periods for these stars, as explained below. For other stars, the sequence has a typical width of $\lesssim$1\,dex.

We also note that cluster members without measured rotation periods follow the same sequences as other cluster members. Hence, these stars apparently follow the very same activity (and, hence, rotational) evolution patterns \citep[cf.][]{2021A&A...656A.103F, 2023A&A...675A.180G}. The lack of a determined rotation period suggests that such stars simply happen to have unfavourable spot configurations or inclination angles.

\subsubsection{Kraft break in the colour-activity diagram}
\label{sec:kraft}
We can clearly detect the transition from cool stars with outer convection zones to intermediate mass stars with primarily radiative envelopes, the so-called Kraft-break \citep{1967ApJ...150..551K}. Blueward of spectral type F5 ($(G-G_\mathrm{RP})_0\approx0.4$), we find a strong drop-off in the activity levels, with stars blueward of the Kraft-break ($\sim$F2) not following a mass-dependent X-ray activity and instead showing a wide spread in the range $-6\lesssim\log L_\mathrm{X}/L_\mathrm{bol}\lesssim -5$. In the rotation period distribution, we see a corresponding flattening of the rotational distribution, again accompanied with a spread of one order of magnitude (Fig.~\ref{fig:CPD} in this work and Fig.~10 in \citealt{2021AJ....162..197B}).

In summary, stars with thin convective envelopes not only have short rotation periods but also have a consistent level of X-ray emission (over the small mass-range observable in NGC\,2516). However, these stars cannot follow the typical rotation-activity relation because, firstly, their convective turnover timescales are very short\footnote{If the relation of \cite{2011ApJ...741...54C} is assumed to be valid to 7000\,K, the convective turnover timescale is on the order of minutes (and is likely within the stellar atmospheres). For other models, such as \cite{2010ApJ...721..675B}, these stars are beyond the range of (interior) models with surface convection zones.} and hence their Rossby numbers are not comparable to lower-mass stars. Secondly, the emission and rotation are constant for a large mass-range in which the convective envelope has varying depth, which would further spread the stars in Rossby number. The constant level of X-ray emission with respect to the bolometric flux is a known fact for earlier type stars \citep[e.g.][]{1981ApJ...248..279P}.

The active A stars apparently form two groups blueward of $(G-G_\mathrm{RP}=0.3)$. With the current number of stars, we cannot be sure if this is a real feature of the CAD or a result of small-number statistics. Yet, in the data presented here, the lower branch appears to be flat with respect to colour at a level of $\log L_\mathrm{X}/L_\mathrm{bol}\approx -5.7$, while the upper branch appears to be rising with stellar mass. We do not speculate here on the possible origin but note that X-ray emission of A stars is not well-understood \citep[see e.g.][]{2022AJ....164....8G}. These stars have also been observed repeatedly by TESS, and a detailed asteroseismic analysis \cite{2024A&A...686A.142L} could hint at potential origins although \cite{1998PASP..110..804Z} have not found any correlation between pulsations and X-ray activity derived from ROSAT data.
NGC\,2516 will also be observed in the Southern PLATO field \citep{2025A&A...694A.313N} for at least two years. These continuous observations might provide further insights in the future.

\section{A closer look at rotation, activity, and mass}
\label{sec:CADrot}

The combination of the large number of X-ray sources assembled from eROSITA and other missions together with the large set of rotation periods from both dedicated and generic time series observations makes the NGC\,2516 data uniquely useful. This situation allows us to better understand the details of the coronal activity CAD and, especially, to explicate the influence of rotation. We first present our analysis broadly in the traditional way to motivate our subsequent detailed subdivisions of the X-ray sources in the sections below.

\subsection{Traditional interpretation of the rotation-activity relationship}

In the standard scheme of interpreting the X-ray rotation activity relationship \citep[e.g.][]{2011ApJ...743...48W}, cool stars are divided into two regions, a saturated one at small Rossby numbers, and an unsaturated region for larger Rossby numbers. The break between these two regions is not (yet) clearly defined for reasons that include:
1. the intrinsic scatter in the X-ray fluxes,
2. the variation in Ro values between different models of the convective turnover timescale, and the largest,
3. the (typically unavoidable) usage of heterogeneous sources for X-ray data, including the necessity of using field stars to assemble large-enough samples to identify trends. We argue for more definitive distinctions through the usage of a single, rich, and revealing open cluster.

\begin{figure}
    \includegraphics[width=\columnwidth]{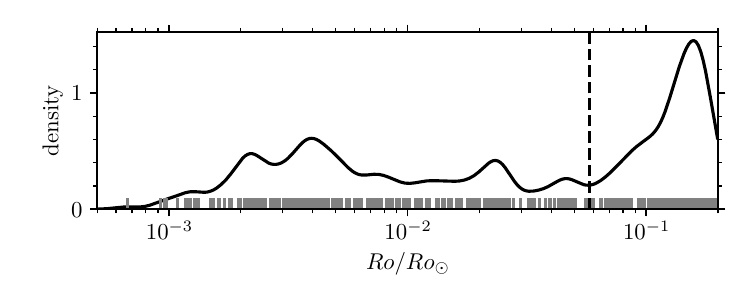}
    \caption{Distribution of solar scaled Rossby number for NGC\,2516 cluster members with rotation periods, together with a traditional break point between saturated and unsaturated stars at $Ro/Ro_{\odot}=0.055$ (vertical dashed line). The grey ticks represent all measured Rossby numbers, while the black line shows a kernel density estimate of their distribution.}
    \label{fig:rohist}
\end{figure}

\begin{figure*}
    \includegraphics[width=\textwidth]{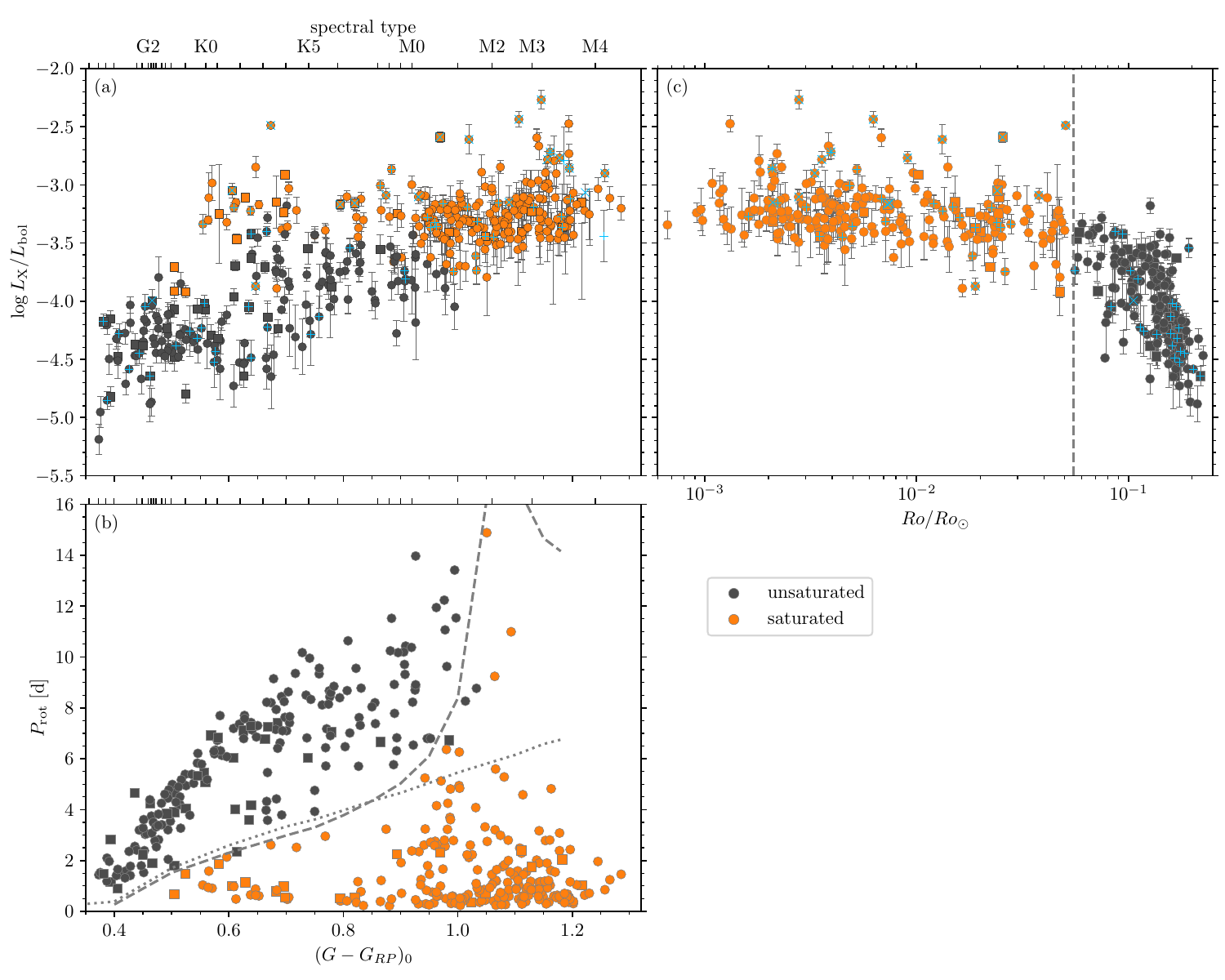}
    \caption{
    Connections between colour, activity, and rotation in the traditional interpretation, wherein stars are classified into two categories, saturated and unsaturated. (Only cluster members with both rotation periods and X-ray detections are plotted in a 3-way layout, allowing individual stars to be cross-identified across panels.)
    Panel (a) (\emph{top left}) shows the colour-activity diagram.
    As in Fig.~\ref{fig:rotact}, members with identified flares in their light curves are marked with blue x-shaped symbols, and members with two potential optical counterparts with blue pluses.
    Panel (b) (\emph{bottom left}): The corresponding colour-period diagram, indicating dividing lines of equal Rossby number separating saturated and unsaturated stars. Lines for $Ro/Ro_\sun=0.055$ are displayed for the \cite{2010ApJ...721..675B} (dashed line) and \cite{2011ApJ...741...54C} (dotted line) prescriptions. The four stars marked unsaturated below the separating line are the result of using a different colour to transform to the convective turnover timescale (see text).
    Panel (c) (\emph{top right}): The corresponding rotation-activity diagram for the cluster members with measured rotation periods. The vertical line separating saturated and unsaturated stars corresponds to the line $Ro/Ro_\sun = 0.055$ in the CPD.
    }
    \label{fig:cornerdiag_simple}
\end{figure*}

The NGC\,2516 X-ray rotation-activity diagram displays several small gaps in the transition region between the saturated and unsaturated regime. To investigate them, we show in Fig.~\ref{fig:rohist} a kernel density estimate (KDE) of the distribution of $\log Ro/Ro_\sun$ using a Gaussian kernel and a bandwidth $h=0.01$. The low Rossby number regime is characterised by several sub-peaks, while the higher Rossby number regime (i.e. unsaturated regime) shows one smooth distribution. The local minimum of the KDE in Fig.~\ref{fig:rohist} falls close to the obvious change in slope of the running mean in Fig.~\ref{fig:rotact} at $Ro/Ro_\odot = 0.04$. Therefore, we use the associated value of Ro, namely $Ro/Ro_\odot = 0.055$, as a natural break-point for the two components of the rotation-activity relationship. If, on the other hand, we were to use the X-ray activity alone (see Fig.~\ref{fig:CAD}) to indicate the break-point, we could argue for a break in the data at $\log L_\mathrm{X}/L_\mathrm{bol} \sim -3.5$ or perhaps even as low as $-3.75$. This would locate the break at $Ro/Ro_\sun \sim 0.1$. Indeed, we could also plausibly make the case for locating the break at an even shorter value of $Ro/Ro_\sun \sim 0.03$. In a very real sense, as we show below, this is an intermediate region that samples a distinct transition between saturated and unsaturated X-ray emission, with the transition being  obvious for early-K stars (or less obvious for later type stars).

This is best seen in the three-way diagram (CPD, CAD, and rotation-activity diagram) shown in Fig.~\ref{fig:cornerdiag_simple}, where we have made a simple two-fold separation at $Ro/Ro_\sun = 0.055$ between saturated and unsaturated stars according to this (objective) criterion. In the CAD seen in Fig.~\ref{fig:cornerdiag_simple}, this division unsurprisingly categorises almost all the M\,dwarfs and the most active K\,dwarfs in the saturated category, while simultaneously placing all the G-type stars and the less active K\,dwarfs in the unsaturated category. This division appears not to be distinctive purely from the X-ray point of view, and merely moving the dividing line to and fro (e.g.  to shorter values of $Ro$) is unlikely to be more convincing. A more nuanced interpretation is offered in Sect. 6.3 below.

Regarding the CPD, we see (unsurprisingly) from the figure that the saturated stars correspond to what have been traditionally roughly labelled the fast rotators in the CPD, while the unsaturated stars correspond to the slower rotators. The separator is the line of constant Rossby number, $Ro/Ro_\sun = 0.055$, as displayed in the figure. (Because of this, there is a mass-dependent distinction between fast- and slow rotators.) However, this separation apparently does not do justice to the CPD. For instance, the slow rotators form a distinctive higher-density sequence in the CPD, both here, and in a series of clusters that have been observed to date (e.g. \cite{2023A&A...672A.159G} and references therein).
However, here in Fig.10, with the $Ro/Ro_\sun = 0.055$ division, the slow rotator sequence (nominally the high-density sequence at the upper edge of the CPD) appears to have been augmented by a set of faster-rotating stars from which it is likely rotationally distinct. (These faster rotators are sparser in the CPD and are in the region named `the rotational gap' by \citealt{2003ApJ...586..464B}.) A similar situation exists for the saturated stars in the CPD, with the faster rotators including not just the sequence of fast rotators (the so-called C-type stars of \citealt{2003ApJ...586..464B}), but also a number of significantly slower rotators, what \cite{2003ApJ...586..464B} called gap stars.

\subsection{Influence of binarity}

To what extent binarity influences stellar activity and rotation directly or not is still an open question. \cite{1994MNRAS.266..798P} found enhanced X-ray luminosities for K-type binaries in Praesepe. The more recent study of \cite{2022ApJ...931...45N}, however, found only minor hints in limited mass ranges in Praesepe and the Hyades which in some cases could also be attributed to a difference in rotation.

As seen from Fig.~\ref{fig:cornerdiag_simple}a (the CAD), almost all the candidate binaries (square symbols) are co-located within the single star X-ray distribution. They do not appear to be preferentially elevated above it. Indeed, certain slow rotator binaries even appear to be under-luminous in X-rays. However, it is true that fast rotators tend to have an elevated binary fraction as compared with the slow rotators. A plausible rationale for this is that stellar binarity could result in earlier dissipation of the circumstellar disc, enabling such stars to retain a greater portion of their angular momenta, as compared with a coeval cluster star that is a slow rotator \citep[e.g.][]{1993AJ....106..372E, 2021A&A...652A..60F}.

This suggests that understanding whether binarity has an influence on the activity would potentially require slicing the data by rotation, mass, and also binarity. Even in our rich sample, the number of stars is inadequate. Therefore, a statistical analysis is not possible for all but the highest mass stars, which of course have already settled on the slow rotator sequence by even the very youthful age of NGC\,2516. For these stars, the differences still appear to be on the order of the uncertainties. Thus, we cannot claim to have detected significant differences in the activity of single and binary stars.

To summarise, even the large dataset presented here does not allow for a meaningful analysis with respect to the X-ray effects of binarity because the influence of differing rotation rates is overwhelmingly dominant over the possible influence of binarity. We note that outliers in the figures throughout this paper consist of both potential binaries and single stars. This suggests that any influence of binarity is camouflaged under the greater variability from other effects.

\subsection{A more nuanced interpretation}
\label{nuanced}
\begin{figure*}
    \includegraphics[width=\textwidth]{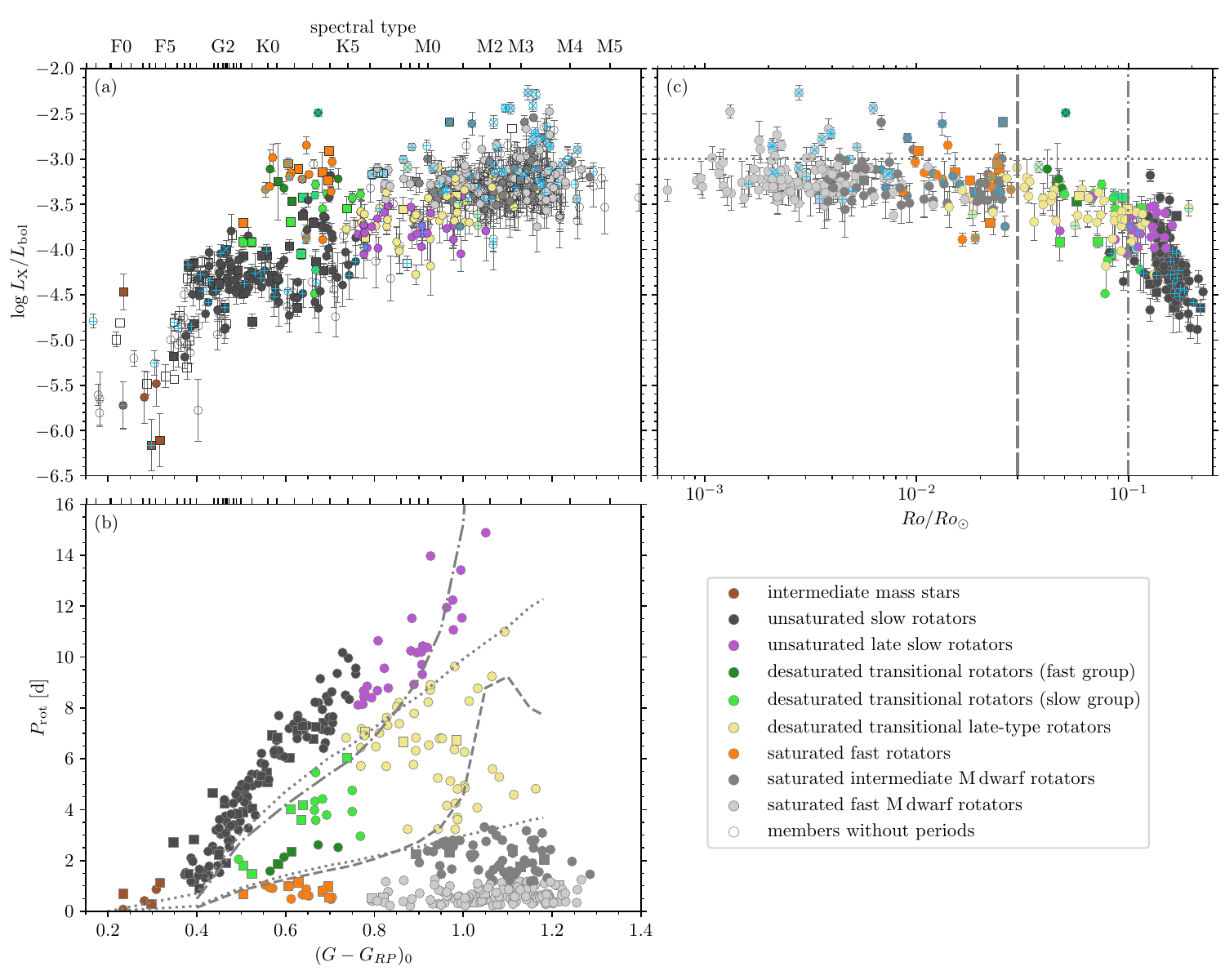}
    \caption{Three-way connections between colour, activity, and rotation, but with more nuanced separations than traditional.
            Panel (a) shows the X-ray colour-activity diagram with stars colour-coded by rotational properties ((\emph{top left}; also see key). As in Fig.~\ref{fig:CAD}, members with identified flares in their light curves are marked with x-shaped symbols, and members with two potential optical counterparts with pluses.
            Panel (b) : Colour-period diagram in which the different rotators are defined (\emph{bottom left}). The lines in the CPD are lines of constant normalised Rossby number, and are located at $Ro/Ro_\sun = 0.03$ and $0.1$, indicating respectively the transition from saturated to desaturated stars, and then to unsaturated stars. A pile-up of rotation periods for early M\,dwarfs just below the line can be seen. The dash-dotted line at longer rotation periods indicates the limit to the unsaturated stars. The two dotted lines show the lines of equal Rossby number at the same $Ro/Ro_\sun$ positions but using $\tau_c$ from \cite{2011ApJ...741...54C}.
            Panel (c) : Rotation-activity diagram for cluster members with their rotation periods available (\emph{top right}). The two vertical lines again correspond to the lines in the CPD, at $Ro/Ro_\sun$ = 0.03 and 0.1.
     }
    \label{fig:cornerdiag}
\end{figure*}

Because our sample is rich in both X-ray detections and rotation periods, we are able to investigate the transition region further, and argue here for a more nuanced appreciation of the transitions (both rotational and X-ray), including understanding the differences in the nature of the transition for different spectral types. The treatment here runs essentially in parallel with that published in \cite{2021A&A...656A.103F}, where similar patterns are seen in the behaviour of calcium IRT activity in the 300\,Myr-old open cluster NGC\,3532. We used a similar colour-coding scheme to that publication, initially based on small groups of stars in the CPD, and subsequently applied it  to all stars in Fig.~\ref{fig:cornerdiag}, so that the correspondences are rendered evident.

We begin with the intermediate mass (F5 and earlier-type) stars (included in Fig.~\ref{fig:cornerdiag}); for reasons discussed earlier, we simply marked these with distinguishable symbols and ignored them in the subsequent cool-star classification. We then divided the remaining cool stars in the CPD into a three-fold classification, roughly (but not exactly) guided by two lines of constant Rossby number at $Ro/Ro_\sun = 0.03$ and $Ro/Ro_\sun = 0.1$. This division categorises the rotating stars into a set of fast rotators that define a high-density region at the bottom edge of the CPD, a set of slow rotators that are also high-density in the CPD (at the upper edge of the rotational distribution) and which appear to define the slow-rotator sequence, along with a group of stars that have intermediate rotation rates that occupy the low density region between them. We note that this last group roughly corresponds to the rotational gap of \cite{2003ApJ...586..464B}. Glancing over at the CAD and rotation-activity diagram indicates that the fast rotators also correspond on a star-by-star basis with a set of stars at the highest levels of X-ray activity, suggesting that they ought to be identified with saturation. Symmetrically, the slow rotators correspond to stars that are both on the low-activity side of the distribution in a mass-dependent way, and occupy the declining region of the rotation-activity relationship, traditionally called 'unsaturated' stars. The stars between these in the CPD (green and yellow symbols in Fig.~\ref{fig:cornerdiag}) have X-ray activity levels that are intermediate between these levels, again in a mass-dependent way. We simply refer to these as 'desaturated' stars. We further divided each of these three categories by mass to facilitate the detailed discussion presented below.

\subsubsection{Rotation in the colour-activity diagram}

From the colour-coded CAD in Fig.~\ref{fig:cornerdiag}a it is immediately obvious that the wide spread in X-ray flux (a factor of 30) among K-type stars is mostly a consequence of the corresponding spread in rotation periods, visible in the CPD in Fig.~\ref{fig:cornerdiag}b. We find that the stars in the CAD are ordered vertically, i.e. stratified by their rotation periods, with the fastest rotators (orange) having the highest X-ray activity, and the intermediate-period stars (green) and slow rotators (black) having successively lower activity levels that could be called desaturated and unsaturated, respectively. This stratification continues for lower-mass (M-type) stars, albeit with greater noise and a smaller X-ray activity range. It is remarkable that K dwarfs have such a large range in activity compared to M dwarfs despite their smaller spread in Rossby number. As seen from the estimated detection limit in Fig.~\ref{fig:CAD}, the X-ray activity distribution is not truncated for the early M\,dwarfs. However, the mid-M regime and the faintest stars in our sample are likely to be affected by the detection limit.

It is curious (and likely meaningful) that there is an exact parallelism between X-ray activity and chromospheric activity behaviours. In comparison with our work on chromospheric activity in NGC\,3532 \citep{2021A&A...656A.103F}, we find these X-ray sequences to be somewhat noisier, but to have an essentially similar structure. In particular, the steep decline in activity when approaching the Kraft break from the cool star side (Sect.~\ref{sec:kraft}) and the spread in activity for K-type stars are comparable. Here, we do not find a branch of decreasing activity with decreasing mass among the M-type stars, the so-called extended slow rotator sequence \citep{2021A&A...656A.103F}. This is simply because we have not been able to measure the X-ray emission for the slowest rotators on the extended slow rotator sequence in NGC\,2516 as they are outside the eROSITA FoV.

An obvious and striking feature of the CAD is that in the same region in which the sequence widens due to the spread in rotation (among early K) some slow rotators seem to have lower $\log L_\mathrm{X}/L_\mathrm{bol}$ values than the higher mass (G-type) slow rotators in the same cluster. We have verified that this drop is real, and not caused e.g. by the bolometric correction used by actually applying the bolometric correction from \cite{2015ApJ...804...64M}. This dip also seems to be unrelated to any possible binarity status of the corresponding stars. We note that a similar dip can also be observed in Praesepe \citep[see Fig.~9 of][]{2022ApJ...931...45N}. This relative decline in X-ray emission for K-type stars immediately after transitioning from fast- to slow rotation therefore appears to be a real feature of the X-ray behaviour of cool stars.

\subsubsection{A gap and a pile-up in the rotation-activity diagram}
Moving on to the rotation-activity diagram in Fig~\ref{fig:cornerdiag}c, we note that within the aforementioned desaturated region two sparsely populated gaps emerge. The first gap can be found near $Ro/Ro_\sun=0.03$ and the second near $Ro/Ro_\sun=0.055$ (cf. local minima of the KDE in Fig.~\ref{fig:rohist}). We can trace both gaps back to the corresponding gaps in the CPD. The latter gap is likely caused by the low population of stars in the centre of the rotational gap and the bimodality of gap stars. Stars apparently simply spin down rapidly through the rotational gap, leaving it relatively unoccupied.

In order to locate the corresponding feature at $Ro/Ro_\sun = 0.03$, we add a line of constant Rossby number, at $Ro/Ro_\sun = 0.03$, to the CPD in Fig.~\ref{fig:cornerdiag}. This line of constant Rossby number divides the CPD into a group of saturated stars below and a group of desaturated and unsaturated stars above. Two facts are notable. Firstly, this line of constant Rossby number splits the evolved fast rotators cleanly from the fast rotators that are still on the flat sequence (near $(G-G_\mathrm{RP})_0=0.5$). Secondly, among the early M dwarfs, we find a pile-up of stars just below that Rossby threshold. The latter feature has been seen before and is found in open clusters of all ages. \cite{2022AJ....164...80R} found a very prominent pile-up of rotators in the very same position for young (16\,Myr) stars in Upper Centaurus–Lupus and Lower Centaurus–Crux. Just above this line of constant Rossby number, a gap can be found in the much older Praesepe.

The persistence of this feature for M\,dwarfs and also the clean division of the fast rotators of higher mass leads us to suggest that $Ro/Ro_\sun=0.03$ defines the onset of the spin-down from fast to slow rotation. We believe that our data justifies the existence of a distinctive intermediate `desaturated' phase in the rotation-activity behaviour of cool stars that lies between the fast and slow sequences. A similar hypothesis was advanced by \cite{2003ApJ...586L.145B, 2003ApJ...586..464B}, where an intermediate 'g' phase was proposed to exist between the fast (C) and slow (I) sequences.

In such a picture, as a star spins down, it moves from left to right in the rotation-activity diagram. Stars that are initially fast rotators spin down with saturated activity until they reach $Ro/Ro_\sun = 0.03$. Immediately below that threshold value,  stars appear to pile up in the CPD, a fact which indicates that the spindown until that point is a relatively leisurely process. Once a star crosses the threshold (of $Ro/Ro_\sun = 0.03$), the spindown appears to become very efficient, and seems to be approximately exponential \citep[c.f.][]{2003ApJ...586..464B, 2023A&A...674A.152F}. Stars seem to cross this desaturated region quickly before settling on the unsaturated slow rotator branch.

The evolutionary picture outlined above explains:
1. the pile-up of stars shortwards of $Ro/Ro_\sun=0.03$,
2. the sparse population of stars between the fast and slow rotator sequence (the gap/desaturated region), and
3. the decline in activity before stars settle on the unsaturated branch.
We note that such a three-segmented explanation has been proposed before in \citet{2003ApJ...586L.145B}, and can now be underpinned with both a larger amount of data, and also within  a single open cluster (meaning a homogeneous setting). This allows us to exclude biases that can arise when combining stars of different ages in the rotation-activity diagram.

\subsection{The rotation activity relationship}

\begin{figure*}
    \includegraphics[width=\textwidth]{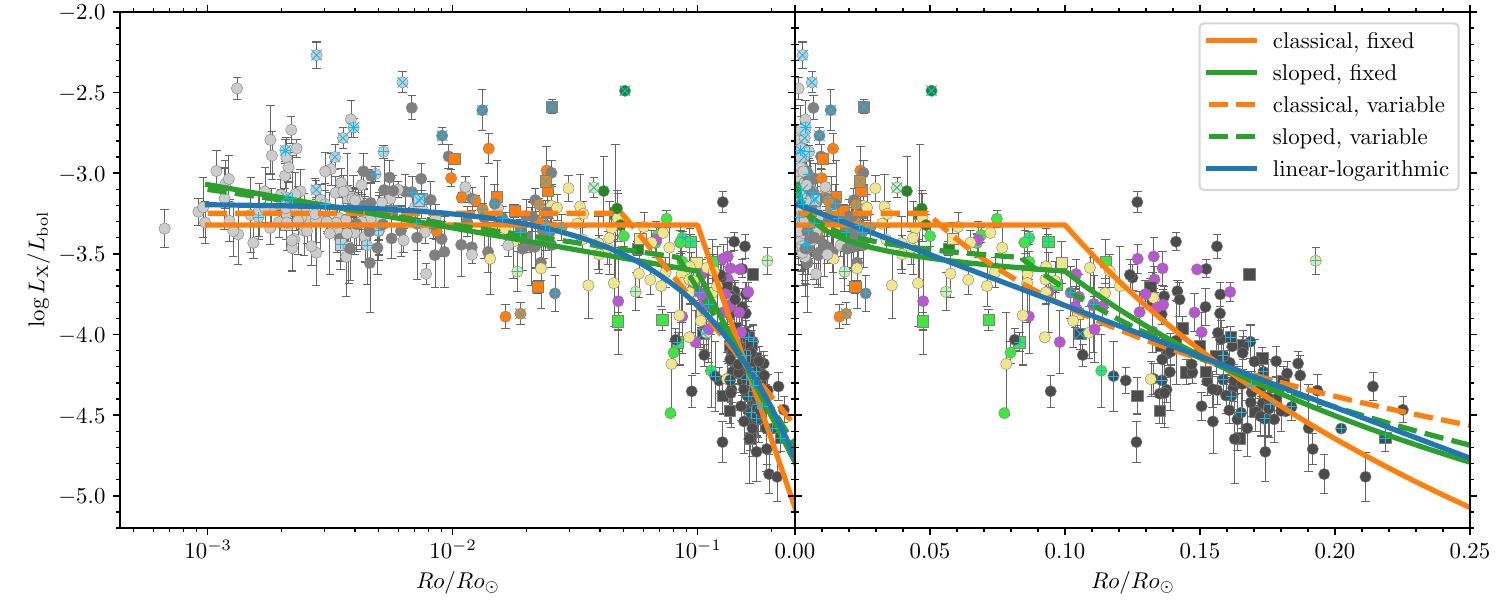}
    \caption{Rotation-activity diagram as in Fig.~\ref{fig:cornerdiag}c, but now displaying different possible models for the rotation-activity relationship. The orange lines show classical segmented power law fits with a constant saturated regime, while green lines show power law fits  with a slope in the saturated regime. The solid (green and orange) lines have a fixed break point at ${Ro/Ro_\sun}_0=0.1$, while it is a free model parameter for the models shown with the dashed lines. The blue line shows the linear-logarithmic model. It is remarkable that a 2-parameter fit can come closest to describing the X-ray behaviour of such a wide range of stars. Both panels show the same content, logarithmic with $Ro/Ro_\sun$ (left) and linear with $Ro/Ro_\sun$ (right). For clarity, we do not show the fits to the unsaturated regime alone.
    }
    \label{fig:rotact_fits}
\end{figure*}

The detailed information about the rotational categories from the previous section enables us to probe the dependence of the fractional X-ray luminosity on the Rossby number for different rotational regimes, going beyond a simple two-way split in Rossby number. In previous work on the rotation-activity relationship, the saturated stars were split off, and the dependence for the unsaturated stars was described with a power-law model (e.g. \citealt{2011ApJ...743...48W}) with a power-law index $\beta\approx-2.5$.

\citealt{2011ApJ...743...48W} worked with a heterogeneous collection of both field stars and those from a number of open clusters.
With the advantage of precisely knowing the rotational state of the stars in NGC\,2516, we can probe the dependence of the power-law index on this state before attempting a fit to the entire distribution in Fig.~\ref{fig:rotact_fits}\footnote{To minimise confusion, we do not display these fits in Fig.~\ref{fig:rotact_fits}.}. For only the spun down unsaturated slow and late-slow rotators (black and purple in the figures), and using a simple $\chi^2$ minimisation we find that $\beta=-1.74\pm{0.12}$, while also excluding flaring stars. Relying only on a cut in Rossby number with $Ro/Ro_\sun>0.1$, we find that $\beta=-2.60\pm{0.12}$. This value indicates a much steeper slope, closer to the typical value in the literature. Nevertheless, we advocate the use, whenever possible, of only the truly slow rotators defined via the colour-period diagram, rather than simply a cut in Rossby number. For this reason, it might not be especially meaningful to compare results from different publications unless only stars in the same rotational regime are compared.

The full distribution is often fitted by a segmented power law. Although the distinction between the slow rotators and stars in transition to slow rotation is not clear cut in Rossby number, we have to adopt a Rossby cut to fit the full distribution.
We fit four different power law models to investigate the dependence of their parameters on the parametrisation and to find the best fitting model. The first class of models uses a fixed break point at ${Ro/Ro_\sun}_0=0.1$, while the second class allows for a variable break point to find the best model.
Each class contains two different parametrisations of the saturated regime, once with a constant saturation value (``classical'', orange in Fig~\ref{fig:rotact_fits}) and one where the saturation level is allowed to change with ${Ro/Ro_\sun}$, namely is treated as a power law itself, with a slope different from the unsaturated regime (``sloped'', green in Fig.~\ref{fig:rotact_fits}) to account for the obvious structure at $Ro/Ro_\sun<0.1$.

To find the best model in this context, we use a least squares fit for each of the four models and compute the Akaike information criterion (AIC) and Bayesian information criterion (BIC). As seen from the results presented in Table~\ref{tab:models}, the first class of models perform worse overall than the second class with a variable break point.
The position of the break point among the variable models is located in the rotational gap, meaning that the derived slope of the power law does not describe the unsaturated regime well, and is hence not a good physical model.
We did not investigate more complex models with three regimes here because of the diminishing value of additional parameters, and because we find an alternative solution to be more appropriate and economical.

\begin{table*}
    \caption{Parameters and information criteria for different models of the activity-rotation relation with the best-fitting model highlighted in bold.}
    \label{tab:models}
    \begin{tabular}{lcccccl}
        \hline
        \hline
        Model & parameters & & & AIC & BIC & style in Fig.~\ref{fig:rotact_fits}\\
        \hline
        &&&&&&\\
        \hline
        Linear-logarithmic & a & b & & & & \\
        \hline
        \textbf{continuous} & $\mathbf{-6.31\pm0.18}$ & $\mathbf{-3.19\pm0.02}$ & & \textbf{-7.49} & \textbf{0.10} & \textbf{solid blue}\\
        &&&&&&\\
        \hline
        Broken power law & $\beta_1$ & $\beta_2$ & ${Ro/Ro_\sun}_0$ & & & \\
        \hline
        classical, fixed & -- & $-4.40$ & $0.1$ & 94.78 & 102.38 & solid orange\\
        sloped, fixed & $-0.27$ & $-2.98$ & 0.1 & 15.72 & 27.11 & solid green\\
        classical, variable & -- & $-1.86$ & $0.05$ & 12.89 & 24.29 & dashed orange\\
        sloped, variable & $-0.22$ & $-2.44$ & $0.08$ & 13.66 & 28.85 & dashed green\\
        \hline

\end{tabular}
\end{table*}

This alternative description of the rotation-activity dependence is a linear-logarithmic form as used in \cite{2008ApJ...687.1264M}: $\log L_\mathrm{X}/L_\mathrm{bol} = a \,Ro + b$. We apply this model to all stars in the sample (blue in Fig.~\ref{fig:rotact_fits}).
It does not require a break point at all.
Moreover, this model describes our data best, a fact also reflected in its much reduced AIC and BIC values.
The preference for this model originates not purely from its lower number of parameters but also because it has the lowest sum of residuals for all considered models.
By virtue of including all measured stars, desaturated and saturated stars are of course included, and distinctions between categories becomes moot. Notably, the y-axis intercept of this model is $a=-3.19\pm{0.02}$, which is very close to the traditional saturation limit of $-3.0$. Therefore, from the point of view of only this cluster, the case can be made that this single fit to all stars is preferential to fitting any alternative sub-groups of stars, especially if consensus does not exist on how such sub-groups are to be created. Whether this viewpoint can be extended to include other clusters will have to be determined in future works.

Our analysis of the rotation-activity diagram suggests the following summation. If we were to follow the traditional prescription of separating saturated and unsaturated stars, then there is no barrier to dividing stars into further categories, for instance, recognising desaturated stars as those undergoing a simultaneous rotational and X-ray transition from fast and saturated to slow and unsaturated. On the other hand, a reasonably persuasive case can also be made for re-integrating such categories back into a single set of stars that can be fit simply with a linear-logarithmic fit that at face value is superior to fits of individual subsets of stars.

\section{Spectral analysis}
\label{sec:specanalysis}

With a large dataset providing spectra for nearly every star included in the eROSITA observations, we can potentially gain unique insights into the distribution of spectral properties of a wide range of cool cluster members. Below, we analyse both the coronal temperature and abundance with respect to the stellar mass, rotation rate, and activity.

\subsection{Coronal temperature}
The origins of coronal heating have been a long-standing issue in stellar astrophysics.
\cite{1960ApJ...132..821P} was able to take a remarkable step forward in predicting and explaining the solar wind by hypothesising a dynamic and outflowing (as opposed to static) solar atmosphere where the corona was assumed to be isothermal, and roughly at a temperature of 1\,MK. Such coronae only became directly observable with the launch of X-ray satellites. Although the energy resolution of today's X-ray satellites allows multi-temperature models to be constructed in certain cases, the weighted mean temperature from two-temperature models are typically reasonable for data \citep{2005ApJ...622..653T} and stars such as ours, and do not run the risk of overfitting that additional temperatures might incur. Given that, we could further ask whether or not such coronal temperatures change as a function of stellar mass, or perhaps with another variable such as the rotation rate (see e.g. \citealt{2003AdSpR..32..937N} in a field star context). Such questions are, of course, better posed in the context of a single open cluster, as we do below, rather than for a field star population, where additional variables such as differing ages and metallicities can cause confusion.

\subsubsection{Distribution in colour}
As outlined in Sect.~\ref{sec:spec_fit}, we modelled each spectrum with a two-component \texttt{APEC} model. Here, we discuss their mean temperatures, weighted by the emission measures of the two components, and we briefly discuss the individual components as well as their relative emission measures in Appendix~\ref{app:twotemp}.
In the left panels of Fig.~\ref{fig:coltemp}, we show the mean temperatures against intrinsic colour for the members of NGC\,2516.
The diagram includes only stars for which we could obtain reasonable two component fits. With 272 stars, this is still the largest such sample in the literature. Overall the mean coronal temperatures for most stars are around 0.6\,keV which corresponds to upwards of 6\,MK, so that our temperatures are in excess of Parker's 1\,MK assumption.

\begin{figure*}
    \includegraphics[width=\textwidth]{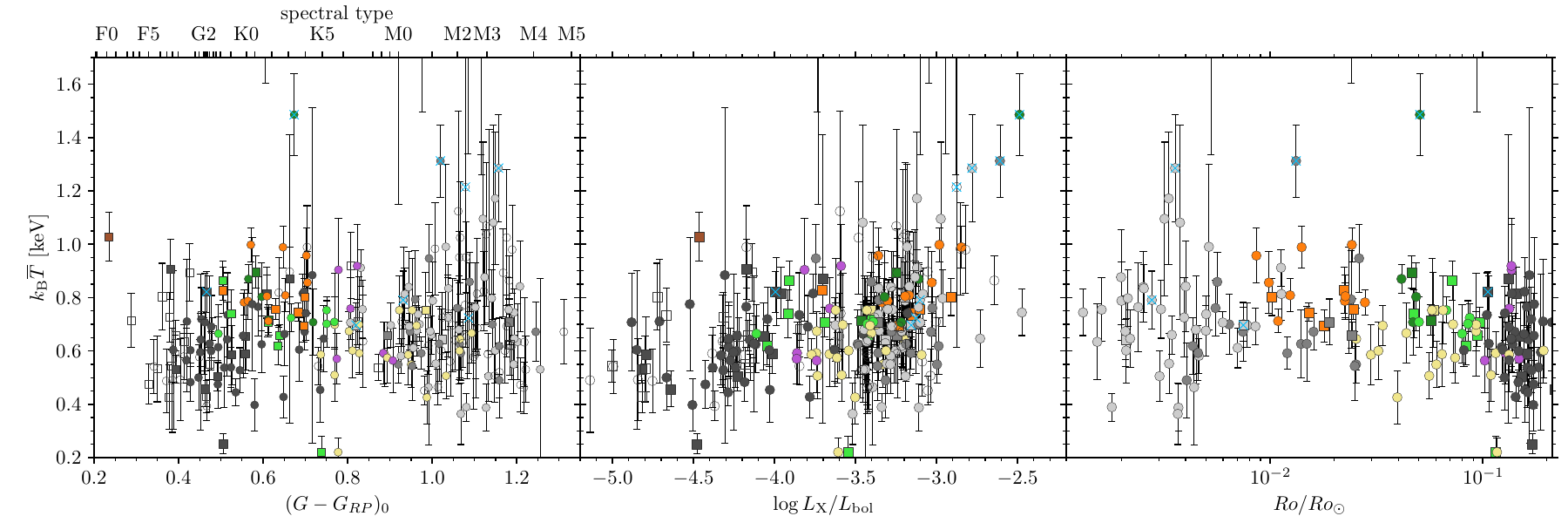}
    \caption{Dependence of the mean coronal temperature on stellar properties.
            We show the quantity $k_\mathrm{B}\overline{T}$ against the intrinsic \emph{Gaia} $(G-G_\mathrm{RP})_0$ colour (\emph{left}), the fractional X-ray luminosity  $\log L_X/L_{\rm bol}$ (\emph{centre}), and the solar-scaled Rossby number (\emph{right}). The colour-coding is the same as in Fig.~\ref{fig:cornerdiag}.
    }
    \label{fig:coltemp}
\end{figure*}

We move from high to low masses to explore the structure of Fig.~\ref{fig:coltemp} in detail. The highest mass stars in our sample (spectral types F and G) show a small mass dependence. The late F stars in the sample have coronal temperatures around $0.4-0.5$\,keV while the later G\,dwarfs have mean temperatures around 0.7\,keV. A notable exception is the bluest star shown in the figure, with a mean temperature above 1\,keV. This star is very active compared to similar stars. We note that it is a binary star in which the secondary component likely contributes to the X-ray flux.

Towards lower-mass stars, we find slightly higher coronal temperatures of $\sim{}0.8$\,keV among early K\,dwarfs. Although this mass regime shows the largest spread in activity, the spread in coronal temperature is small compared to the entire sample. Towards later-type stars the dispersion increases, likely due to the large number of flaring stars. Notably, the coronal temperatures of the non-flaring M\,dwarfs are lower than that of the K\,dwarfs at $\sim{}0.6\,keV$. Flaring stars have mean temperature above 1\,keV and sometimes outside the plotted range in Fig.~\ref{fig:coltemp}.

\subsubsection{Dependence on fractional X-ray luminosity}

In the middle panel of Fig.~\ref{fig:coltemp}, we show the dependence of the coronal temperature on the fractional X-ray luminosity. The more active stars seem to have higher coronal temperatures, although this trend is obscured by the large scatter and uncertainties on our data. In contrast, \cite{2022ApJ...931...45N} found a clear trend of higher temperatures with higher fractional X-ray luminosity.

\subsubsection{Influence of rotation on the coronal temperature}

\begin{figure*}
    \includegraphics[width=\textwidth]{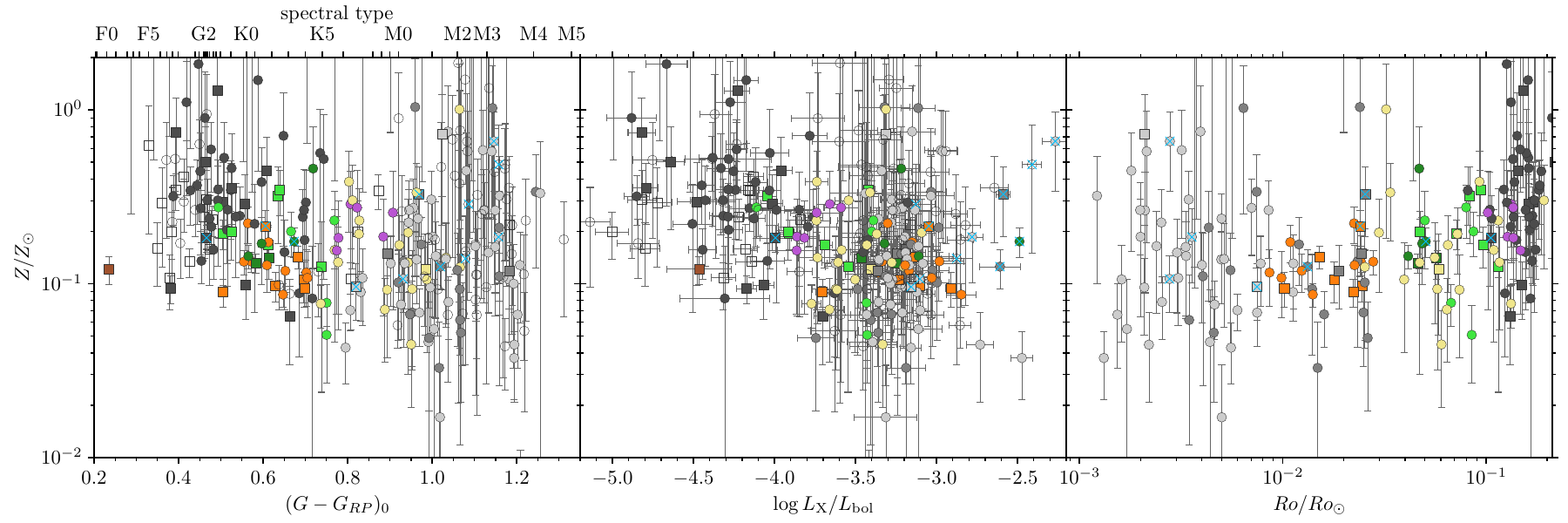}
    \caption{Coronal abundances against three independent variables, similar to Fig.~\ref{fig:coltemp}.
    \emph{Left:} Plotting abundances against intrinsic \emph{Gaia} $(G-G_\mathrm{RP})_0$ colour, we find marginally depressed abundances for late-K and early-M stars, although with a large scatter. Colours and symbols are as in Fig.~\ref{fig:cornerdiag}.
    \emph{Centre:} Plotting against $L_X/L_{\rm bol}$ suggests that the highest coronal abundances are found among inactive stars.
    \emph{Right:} Plotting abundances against solar-scaled Rossby number ($Ro/Ro_\sun$) indicates that marginally higher coronal abundances are found for stars with the highest and lowest values of $Ro/Ro_\sun$.}
    \label{fig:colabund}
\end{figure*}

As in other figures throughout this work, we mark different groups of rotators in Fig.~\ref{fig:coltemp} by colour. We would ordinarily expect the biggest influence of rotation to be found among stars with the largest spread in Rossby number. For our sample these are late G-type and early K-type stars, which exist both as fast and slow rotators (c.f. Sect.~\ref{sec:CADrot}).

Focusing on this mass regime ($0.5\gtrsim(G-G_\mathrm{RP})_0\gtrsim0.75$), we do not find a clear stratification of the different rotation classes as is seen in the colour-activity diagram of Fig.~\ref{fig:cornerdiag}a. The fast and desaturated rotators among the K\,dwarfs have a higher coronal temperature. However, several slow rotators show similar coronal temperatures, not allowing for a definitive conclusion.

In the right-most panel of Fig.~\ref{fig:coltemp}, we show the coronal temperature against the solar-scaled Rossby number. Again, we are unable to find strong evidence of a dependence of the coronal temperature on the rotational properties. However, we note that the fast rotating G and K\,dwarfs (orange) might have a slightly higher coronal temperature. The large scatter in temperatures among the unsaturated slow rotators can be traced back to both the large scatter among early G\,stars and the mass dependence of the coronal temperature for G and K stars.

We find it somewhat surprising that we are unable to identify a clear imprint of the rotational state and hence the activity level on the coronal temperature in NGC\,2516. This might be a consequence of our sample lacking low-activity stars, given that our entire sample consists of young, highly active stars.
In contrast, other work on (small) samples of field stars \citep[e.g.][]{1999ApJ...512..874S, 2005ApJ...622..653T} find lower coronal temperatures for less active stars.
With our dataset, we are not in a position to disentangle the mass-dependence from influences of the rotation and activity. Given the strong mass-dependence of the coronal temperatures, we urge control of this parameter in samples of field stars.

\subsection{Abundances}

The other free parameter in our spectral model, apart from the two temperatures, is the coronal metal abundance.
From solar and stellar observations it is well established that this abundance is different from the photospheric abundance because of the first ionisation potential (FIP) effect \citep{1985ApJS...57..173M, 1995ApJ...443..416L, 2015LRSP...12....2L}.
Inactive stars follow a correlation with stellar mass such that solar-mass stars exhibit the FIP effect with an enhanced coronal abundance, while less massive stars, in particular M\,dwarfs show an inverse FIP (iFIP) effect with a depletion in the corona \citep{2001A&A...365L.324B, 2010ApJ...717.1279W}.
Active stars and evolved cool stars show mostly an iFIP effect \citep{1994ApJ...436L..83A,2022A&A...659A...3S}.

With the current sample of \ndetmem{} members (of which 272 have a well defined \texttt{APEC} model fit and are analysed here), we are able to probe the coronal metal abundances over a large mass-range in a coeval and chemically homogenous population and also to investigate the influence of rotation. Fig.~\ref{fig:colabund} shows the coronal abundances as determined from the \texttt{APEC} model against three independent variables, namely the intrinsic colour, the fractional X-ray luminosity, and the solar-scaled Rossby number. (This plot is comparable with Fig.\,6 in \cite{2022ApJ...931...45N}.) We also highlight the different rotational groups of Fig.~\ref{fig:cornerdiag} to enable identification of any underlying trends. For the whole sample of cluster members, we find a median abundance of $Z=0.21^{+0.46}_{-0.12}\,Z_\sun$ (uncertainties based on the 16th and 84th percentile respectively) for the whole sample of cluster stars. This is similar to the previously assumed value of $Z=0.3\,Z_\sun$ \citep{2006A&A...450..993P}.

In the left-most panel of Fig.~\ref{fig:colabund}, we show the mass dependence of the abundances. We find a trend of lower abundance for lower masses, similar to the one seen by \cite{2022ApJ...931...45N} in Hyades and Praesepe. A similar trend with lower abundances with increasing fractional X-ray luminosity is visible in the central panel. As intrinsic colour and fractional X-ray luminosity are correlated (s. Sect.~\ref{sec:CAD}), we are not in a position to assign fractional responsibilities to the two variables, in particular as both variables have an impact on the strength of the iFIP effect \citep{2022A&A...659A...3S}.

Two groups of stars with exceptionally high abundances stand out in our sample. The first group consists of the most massive (and least active) stars in our sample have abundances close to the Solar photospheric value or even above. These results are in agreement with the higher abundances measured for such stars in \cite{2010ApJ...717.1279W} and \cite{2022A&A...659A...3S}. The second group consists of early M\,dwarfs, which we would not expect to have such high coronal abundances. From both the centre and right panel of Fig.~\ref{fig:colabund}, we find no correlation of these stars with activity or rotation. For every star with an increased abundance, we find a similar star with a normal (low) abundance, suggesting that we have a large scatter in our measurements, as shown by the uncertainties in Fig.~\ref{fig:colabund}.
Flaring stars are distributed evenly throughout the sample, leading us to conclude that the higher abundances are not caused by flaring events.
\cite{2022ApJ...931...45N} apparently did not find such M\,stars in their analysis of Praesepe and the Hyades, despite the higher-mass slow rotators in their sample exhibiting a similar pattern as seen in NGC\,2516.

The right panel of Fig.~\ref{fig:colabund} shows the dependence of the coronal abundance with rotation. We find trends similar to those seen before, and can confirm that the least active stars display higher abundances.

\section{Conclusions}

Here, we present a detailed X-ray study of the young open cluster NGC\,2516 based on the calibration data obtained with eROSITA. Thanks to the sensitivity and large FoV of the telescope, this dataset is one of the richest extant X-ray observations of any open cluster. Together with recently published large sets of rotation periods for stars in NGC\,2516, it enables us to study the dependence of X-ray properties on mass and stellar rotation in previously unseen detail.

In total, we identified \ndet{} X-ray sources within the FoV of the pointed eROSITA observations from the calibration and performance verification programme.
Among these detections, \ndetmem{} stars are cluster members of NGC\,2516, of which \ndetmemrot{} have measured rotation periods.

The rotation activity diagram exhibits the well-known shape of a flat saturated plateau and an unsaturated sequence where the fractional X-ray luminosity decreases steadily with the Rossby number. However, between these two well-known regions, we find a regime of slower decline of activity with Rossby number. This desaturated regime concerns stars in the rotational gap of NGC\,2516 with spectral types K and early-M,  suggesting that the rotation-activity relation is a three-segmented function of Rossby number -- or it might even  turn out to be more structured or complex.

The colour-activity diagram (CAD) constructed from these observations shows, as expected, a trend of lower activity for the warmer stars and higher activity for the cooler stars. There is an intermediate region, highlighted by K-type stars, where the dependence of X-ray behaviour is more complex.
Here, the fastest rotators are among the brightest X-ray emitters of our sample, while the slow rotators have even lower X-ray emission than the warmer G-type stars. Stars that are in transition from the fast-to-slow sequence are intermediate X-ray emitters.

This combination of rotation, colour, and X-ray activity allows us to probe the rotation activity diagram in greater detail than ever before. We identified a gap between the saturated and desaturated regime, which we interpret as the onset of the spin-down from fast to slow rotation. Faster rotators tend to pile up in the CPD and above this Rossby number, fewer stars can be found in the CPD.

Prior work has employed a power-law to describe the correlated unsaturated regime. Using this approach, we find a slope in good agreement with such work. However, we also find that a linear-logarithmic function can not only describe this regime slightly better, but is able to characterise the desaturated regime as well and with a smaller number of parameters.

Our set of two-component coronal temperatures appears to be the largest extant set for an open cluster in the literature. The mean temperatures have slight dependences on stellar mass and relative X-ray luminosity. However, we did not find a clear dependence on the rotation rate.

The coronal abundances are affected by the iFIP effect, and are largely below the solar value, as seen by prior investigators of other clusters. We found a correlation between the abundance and stellar activity, with slow rotators exhibiting higher abundances, occasionally even surpassing the solar value. Other stars with high abundances are probably affected by flaring events.

In parallel with our work on NGC\,2516, \cite{2025A&A...699A.251Y} recently argued for a more nuanced interpretation of the rotation-activity relation of field stars. Based on chromospheric activity indicators measured from LAMOST spectra, they determined a multiply broken relationship of chromospheric emission with Rossby number that has distinct field star parallels with our cluster-based work.

This unprecedented dataset of stellar X-ray observations from the eROSITA CalPV observations has enabled our measurement of the coronal properties of a large sample of stars in NGC\,2516. In combination with recently derived rotation periods, we have gained insights into the dependence of these properties not only on mass, but also on rotation rate. In particular, we show that the rotational transition from fast-to-slow rotation is mirrored in X-rays by stars transitioning from the saturated to the unsaturated regime through a desaturated region that is most clearly defined by K-type stars from our NGC\,2516 sample. Our work affirms the benchmark status of NGC\,2516 for cool star X-ray emission.

\section*{Data availability}
The full Tables 2, 3, 4, and A1 are only available in electronic form at the CDS via anonymous ftp to cdsarc.u-strasbg.fr (130.79.128.5) or via http://cdsweb.u-strasbg.fr/cgi-bin/qcat?J/A+A/.

\begin{acknowledgements}
    We are grateful to the anonymous referee for the helpful report.
    DJF acknowledges support from the Flemish Government under the long-term structural Methusalem funding program by means of the project SOUL: Stellar evolution in full glory, grant METH/24/012 at KU Leuven.
    SAB gratefully acknowledges support from the German Science Foundation (DFG) via BA-4535-1.
    SO is supported by the TUBITAK 2219-International Postdoctoral Research Fellowship Program for Turkish Citizens.
    This work was supported by the Deutsches Zentrum f\"ur Luft- und Raumfahrt under contract numbers 50 QR 2104 and 50 QR 2504.
    This research has made use of NASA's Astrophysics Data System Bibliographic Services.
    This research has made use of the SIMBAD database and the VizieR catalogue access tool, operated at CDS, Strasbourg, France.
    This research has made use of NASA's Astrophysics Data System Bibliographic Services.
    This work is based on data from eROSITA, the soft X-ray instrument aboard SRG, a joint Russian-German science mission supported by the Russian Space Agency (Roskosmos), in the interests of the Russian Academy of Sciences represented by its Space Research Institute (IKI), and the Deutsches Zentrum für Luft- und Raumfahrt (DLR). The SRG spacecraft was built by Lavochkin Association (NPOL) and its subcontractors, and is operated by NPOL with support from the Max Planck Institute for Extraterrestrial Physics (MPE). The development and construction of the eROSITA X-ray instrument was led by MPE, with contributions from the Dr. Karl Remeis Observatory Bamberg \& ECAP (FAU Erlangen-Nuernberg), the University of Hamburg Observatory, the Leibniz Institute for Astrophysics Potsdam (AIP), and the Institute for Astronomy and Astrophysics of the University of Tübingen, with the support of DLR and the Max Planck Society. The Argelander Institute for Astronomy of the University of Bonn and the Ludwig Maximilians Universität Munich also participated in the science preparation for eROSITA.
    This work has made use of data from the European Space Agency (ESA) mission \emph{Gaia} (\url{https://www.cosmos.esa.int/gaia}), processed by the \emph{Gaia} Data Processing and Analysis Consortium (DPAC, \url{https://www.cosmos.esa.int/web/gaia/dpac/consortium}). Funding for the DPAC has been provided by national institutions, in particular the institutions participating in the \emph{Gaia} Multilateral Agreement.
    \newline
    \textbf{Software:}
    The eROSITA data shown here were processed using the eSASS software system developed by the German eROSITA consortium.
    This research made use of \textsc{Astropy}, a community-developed core Python package for Astronomy \citep{2013A&A...558A..33A}.
    This work made use of \textsc{Topcat} \citep{2005ASPC..347...29T}.
    This research made use of the following \textsc{Python} packages:
    \textsc{Stingray} v1.0 \citep{2019ApJ...881...39H,Huppenkothen2019,matteo_bachetti_2022_6394742};
    \textsc{Pandas} \citep{pandas};
    \textsc{NumPy} \citep{numpy};
    \textsc{MatPlotLib} \citep{Hunter:2007};
    \textsc{IPython} \citep{ipython};
    \textsc{SciPy} \citep{scipy};
    \textsc{seaborn} \citep{Waskom2021}

\end{acknowledgements}

\bibliographystyle{aa} % style aa.bst
\bibliography{eRositaN2516.bib} % your references Yourfile.bib

\begin{appendix}

\section{Estimating spurious matches}
\label{app:matches}

The filtered \emph{Gaia} catalogue used in Sect.~\ref{sec:counterp} has a high source density, hence we would expect a certain number of false positive matches.
The eROSITA sources as not evenly distributed in the FoV due to the lower off-axis efficiency and the distribution of stars in the open cluster. To estimate the number of spurious matches among our optical counterparts, we calculated the \emph{Gaia} source density around the eROSITA sources. We created a HEALPix level 12 map of both catalogues and calculated the local \emph{Gaia} source density $\rho_{Gaia}$ in each HEALPix pixel of the eROSITA observations. For each pixel, we calculate the probability of spurious matches through $p=\rho_{Gaia}*A_\mathrm{match}$, with  $A_\mathrm{match}$ our matching area with radius 7\arcsec. The number of expected spurious matches $\sum p$ is 87 sources for the two catalogues. Given that we expect most of the cluster members to be active stars, we are not concerned about the possible false positive matches.

\section{Power-law models of X-ray spectra}
As part of the spectral fitting, we also carried out power-law fits to the spectra. As these are mostly relevant to non-stellar objects, we publish the results in Table~\ref{tab:powerlaw} only for X-ray detected sources without a stellar counterpart in Sect.~\ref{sec:match}.

\begin{table}
    \caption{Key to the online Table of the measured spectral properties for sources without stellar counterparts.}
    \label{tab:powerlaw}
    \resizebox{\columnwidth}{!}{
        \begin{tabular}{lll}
            \hline
            \hline
            Name & Unit & Description\\
            \hline
            SRC\_ID & - & Identifier\\
            RA & $\deg$ & Right ascension\\
            Dec & $\deg$ & Declination\\
            NH & $10^{22}$\,cm$^{-2}$ & Hydrogen column density\\
            PhoIndex & -- & Power-law index of spectral fit\\
            PhoIndex\_err\_l & -- & Lower bound of power-law index\\
            PhoIndex\_err\_u & -- & Upper bound of power-law index\\
            PhoIndex\_norm & -- & Normalisation\\
            Probability & - & Null hypothesis probability of the fit \\
            \hline
        \end{tabular}
    }
\end{table}

\section{The choice of the convective turnover time}
\label{app:tau}
\begin{figure}
        \includegraphics[width=\columnwidth]{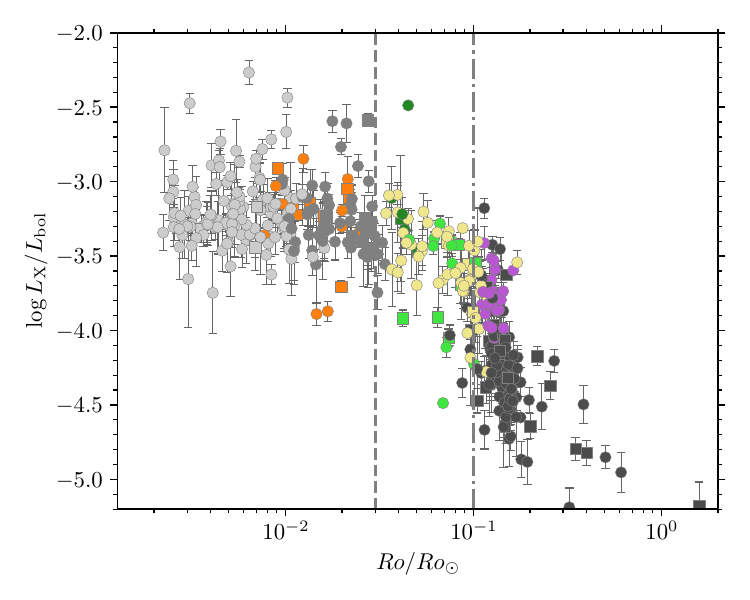}
        \caption{
        Rotation-activity diagram as in Fig.~\ref{fig:cornerdiag}, but here using the convective turnover timescale from \cite{2011ApJ...741...54C}. The vertical lines at $Ro/Ro_\sun = 0.03$ and $0.1$ again represent the transition values from saturated to desaturated and desaturated to unsaturated stars, respectively. This is a consequence and a benefit of the usage of the rescaled Rossby number.
        }
        \label{fig:rotactCS}
    \end{figure}

Several different prescriptions are available for the convective turnover timescale, a quantity that is not directly observable, and which must be calculated. For our work, we adopted the values of \cite{2010ApJ...721..675B}. In order to investigate the influence of our choice on the obtained results, we show in Fig.~\ref{fig:rotactCS} the rotation activity diagram based on the convective turnover timescale from \cite{2011ApJ...741...54C}. This parametrisation mostly differs from \cite{2010ApJ...721..675B} by a factor of $2.7$, except for for M-type stars, where the difference is greater. Despite this, the rotation-activity diagram appears to be remarkably similar to the one described in the main text.

As seen in Fig.~\ref{fig:rotactCS} the overall shape of the rotation activity relation stays the same. Crucially for our conclusions, the transition from saturated to desaturated stars can also be located. Here, we find it at $Ro_\mathrm{CS11}=0.042$.

\section{Two component \texttt{APEC} models}
\label{app:twotemp}

\begin{figure}
    \includegraphics[width=\columnwidth]{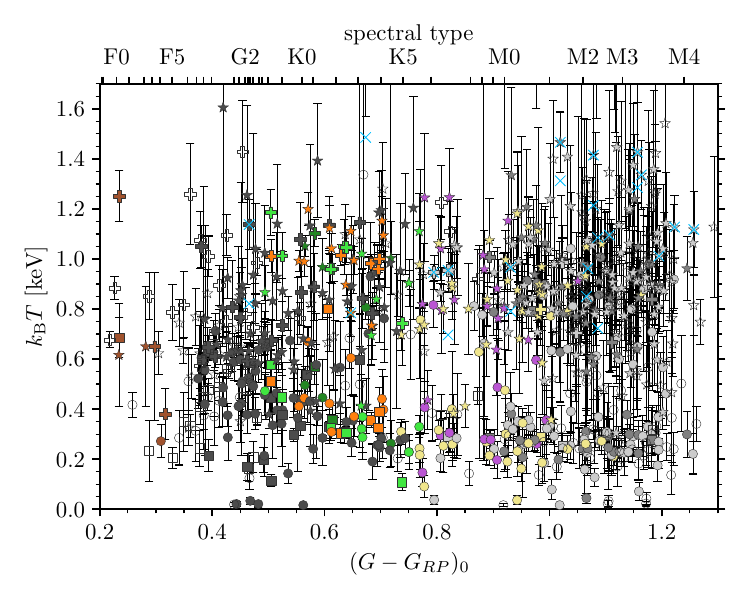}
    \caption{Dependence of the two \texttt{APEC} temperature components on the \emph{Gaia} $(G-G_\mathrm{RP})_0$ colour. Circles and squares denote the first of the two components of the \texttt{APEC} model for single and binary stars, respectively. The asterisks and pluses denote the same two categories for the second component. The colour-coding is the same as in Fig.~\ref{fig:cornerdiag}.}
    \label{fig:coltemp2}
\end{figure}

In Section~\ref{sec:specanalysis}, we analyse the mean coronal temperature based our two component \texttt{APEC} model. Here, we provide a brief analysis of the individual components and their dependence on stellar mass. Fig.~\ref{fig:coltemp2} shows the two components against the \emph{Gaia} $(G-G_\mathrm{RP})_0$ colour. The warm and cool components of most stars except for the warmest ($(G-G_{\rm RP})_0 \sim 0.4$) are distinct, and separated by about 0.7\,keV. The warm components are roughly uniformly distributed, around 1 keV, although with a large scatter. The cool components are mostly in the 0.2-0.5\,keV range. They appear to show a small mass dependence, with G stars having higher temperatures.

While the highest mass (F-type) stars in our sample have an increasing colder component with decreasing mass, we cannot find any coherent structure in their warm component. The increase in the colder component continues to (early) G stars  and we find the highest temperatures (in the cold component) of all stars among these (up to $k_BT_1\approx 0.65$\,keV at $(G-G_\mathrm{RP})_0\approx0.45$). From Fig.~\ref{fig:coltemp}, it appears as if both temperature components are similar for these stars. However, in the same mass-regime, we find a very large spread in temperature in both components. Hence, the overlap observed in the left panels of Fig.~\ref{fig:coltemp} is the result of both the mass independence of the warmer component and the large spread in temperatures. Due to this spread, some stars have their colder component comparable to the warmer component of other stars.

Towards lower-mass stars the colder component decreases in energy up to $(G-G_\mathrm{RP})_0\approx0.7$ and stays constant thereafter. The warm component is mostly mass independent so that two clear sequences emerge with $k_\mathrm{B}T_1\sim 0.25$\,keV and $k_\mathrm{B}T_2\sim 0.9$\,keV. In particular for M\,stars, we find several stars with very hot coronal temperatures most of them effected by flares.

\begin{figure}
    \includegraphics[width=\columnwidth]{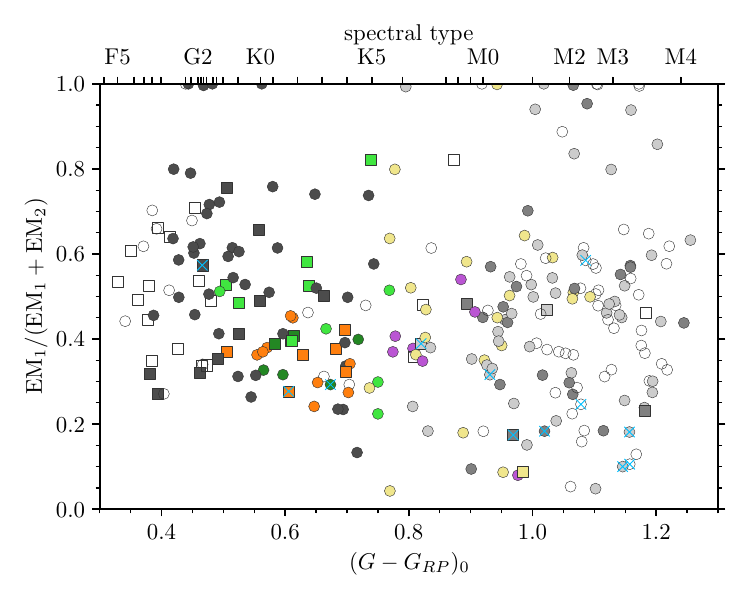}
    \caption{Relative weights of the emission measure against \emph{Gaia} $(G-G_\mathrm{RP})_0$ colour. Most stars have equal contributions of both components in their \texttt{APEC} model. The colour-coding is the same as in Fig.~\ref{fig:cornerdiag}.}
    \label{fig:EM_weights}
\end{figure}

The structure seen in Fig.~\ref{fig:coltemp2} is in agreement with observations by \cite{1990ApJ...365..704S}, who found that F and G-type field stars have mostly only a single temperature component while later type stars have to be described with two components. Further support for this conclusion comes from the relative emission measures as shown in Fig.~\ref{fig:EM_weights}. The slowly rotating G-type stars tend to have a larger weight in the first component, while the bulk of the stars is described by two equal components. Only the very active (and often flaring) stars have a stronger weight on their higher temperature component.

\end{appendix}

\end{document}